\documentclass[english,superscriptaddress]{revtex4-1}
\usepackage[T1]{fontenc}
\usepackage[latin9]{inputenc}
\usepackage{color}
\usepackage{babel}
\usepackage{verbatim}
\usepackage{amsmath}
\usepackage{amssymb}
\usepackage{graphicx}
\usepackage[unicode=true,pdfusetitle,
 bookmarks=true,bookmarksnumbered=false,bookmarksopen=false,
 breaklinks=false,pdfborder={0 0 1},backref=false,colorlinks=true]
 {hyperref}
\usepackage{placeins}
\makeatletter

\providecommand{\tabularnewline}{\\}

\allowdisplaybreaks

\makeatother

\begin{document}
\title{A new formalism for spin alignment with off-shell effects}
\author{Wen-Bo Dong}
\email{wenba@ustc.edu.cn}
\affiliation{Department of Modern Physics and Anhui Center for Fundamental Sciences
in Theoretical Physics, University of Science and Technology of China,
Hefei, Anhui 230026, China}
\affiliation{Institut f\"{u}r Theoretische Physik, Johann Wolfgang Goethe-Universit\"{a}t, Max-von-Laue-Str.~1, D-60438 Frankfurt am Main, Germany}
\author{Yi-Liang Yin}
\email{yinyiliang@mail.ustc.edu.cn}
\affiliation{Department of Modern Physics and Anhui Center for Fundamental Sciences
in Theoretical Physics, University of Science and Technology of China,
Hefei, Anhui 230026, China}
\affiliation{Institut f\"{u}r Theoretische Physik, Johann Wolfgang Goethe-Universit\"{a}t, Max-von-Laue-Str.~1, D-60438 Frankfurt am Main, Germany}
\author{Qun Wang}
\email{qunwang@ustc.edu.cn}
\affiliation{Department of Modern Physics and Anhui Center for Fundamental Sciences
in Theoretical Physics, University of Science and Technology of China,
Hefei, Anhui 230026, China}
\affiliation{School of Mechanics and Physics, Anhui University of Science and Technology,
Huainan,Anhui 232001, China}
\begin{abstract}
We develop a new formalism for studying spin alignment, which can include the off-shell correction and the $\mathcal{O}(\partial)$ contribution in a self-consistent way.
This formalism is based on the Dyson-Schwinger equations expanded to the leading and next-to-leading orders of space-time gradient expansion on a closed time path contour. 
In addition to the well-studied Poisson bracket term, we systematically include the less-explored intrinsic gradient corrections to self-energies. 
Instead of reducing these equations into spin Boltzmann equations with the quasi-particle approximation, we solve them directly and obtain the formal solutions order by order in gradient expansion. 
Applying this formalism to the $\rho$ meson in a pion gas, we find that the spin alignment can arise from the difference between the longitudinal and transverse spectral components at leading order and from thermal shear at the next-to-leading order. 
Although Poisson bracket and next-to-leading order self-energy generate the non-vanishing thermal vorticity contribution individually, these contributions cancel precisely when combined, leading to no contribution to spin alignment induced at first order in thermal vorticity. 
We also make a numerical calculation and our numerical results suggest that the thermal shear may induce a significant spin alignment.
\end{abstract}
\maketitle

\section{introduction}

In non-central heavy-ion collisions, the orbital angular momentum of colliding nuclei can be partially transferred into the medium, leading to the spin polarization of final state particles ~\citep{Liang:2004ph,Liang:2004xn,Voloshin:2004ha,Betz:2007kg,Becattini:2007sr,Gao:2007bc}. 
This phenomenon is called the global polarization effect and has been studied extensively, see ~\citep{Wang:2017jpl,Florkowski:2018fap,Huang:2020dtn,Gao:2020lxh,Gao:2020vbh,Liu:2020ymh,Becattini:2022zvf,Hidaka:2022dmn,Becattini:2024uha} for recent reviews. 
The spin polarization of a vector meson can be described by a $3\times3$ Hermitian matrix, $\rho_{\lambda_{1}\lambda_{2}}$ with unit trace, called the spin density matrix, where $\lambda_{1,2}=-1,0,1$ denotes the spin states along the spin quantization direction. 
There are eight independent variables in $\rho_{\lambda_{1}\lambda_{2}}$, three of them represent the vector polarization and the remaining five represent the tensor polarization. 
In parity conserved two-body decays of the vector meson, only tensor components can be measured through the angular distribution of decay products and the $\rho_{00}$ component is the most commonly measured observable~\citep{Schilling:1969um,Tang:2018qtu}. 
If $\rho_{00}\neq1/3$, the spin states of the vector meson are not equally populated, which is called the spin alignment. 
In heavy-ion collisions the global spin alignment was proposed with respect to the reaction plane~\citep{Liang:2004xn}. 
The STAR collaboration made the first attempt in 2007 to measure the global spin alignment of $\phi$ and $K^{*0}$ mesons in Au+Au collisions at $\sqrt{s_{NN}}=$200 GeV but found no significant signals~\citep{STAR:2008lcm}. 
In 2022, a large positive deviation of $\rho_{00}$ from 1/3 for $\phi$ mesons was observed in the STAR experiment in Au+Au collisions at lower energies while the results for $K^{*0}$ are consistent with 1/3~\citep{STAR:2022fan}. 
The ALICE collaboration measured the spin alignment of $J/\Psi$ in Pb+Pb collisions at $\sqrt{s_{NN}}=5.02$ TeV and a negative deviation of $\rho_{00}$ from 1/3 was reported~\citep{ALICE:2022dyy}. A similar negative deviation was observed in the preliminary STAR results from isobar collisions~\citep{Shen:2024rdq}.

Various theoretical mechanisms of vector meson's spin alignment have been proposed to explain the experimental data~\citep{Sheng:2019kmk,Sheng:2020ghv,Sheng:2022ffb,Sheng:2022wsy,Sheng:2023urn,Li:2022vmb,Dong:2023cng,Muller:2021hpe,Kumar:2023ghs,Fang:2023bbw,Sheng:2024kgg,Dong:2026uvl}. 
In particular, a model based on the fluctuation/correlation in strong force fields~\citep{Sheng:2019kmk} can account for the large deviation of $\rho_{00}$ from 1/3 observed for $\phi$ mesons~\citep{Sheng:2022wsy,Sheng:2022ffb,Sheng:2023urn}. 
For recent reviews on the spin alignment of vector mesons, we refer the readers to, for example, Refs.~\citep{Wang:2023fvy,Chen:2023hnb,Chen:2024afy}.

Recently, the thermal shear contribution to the spin alignment of vector mesons has been investigated using linear response theory~\citep{Li:2022vmb,Dong:2023cng}. 
However, in our previous calculation, the interaction correction to the energy-momentum tensor is neglected because it leads to a divergence problem. 
Consequently, the previous calculation captures only part of full next-to-leading order (NLO) corrections. 
To systematically investigate additional NLO contributions, we develop a formalism based on the direct solution of Dyson-Schwinger equation (DSE) on the closed-time path (CTP). This approach allows us to avoid the quasi-particle approximation and include off-shell and NLO self-energy effects in a self-consistent way. 
We further find that the intrinsic gradient correction to the self-energies can generate a non-vanishing NLO correction which can induce a non-trivial contribution to spin alignment but less discussed. 
For illustration, we consider the $\rho$ meson in a $\pi$ gas with the $\rho$-$\pi$ interaction described by the chiral perturbation theory~\citep{Gale:1990pn,Klingl:1996by,Scherer:2002tk}. Using locally equilibrated pion distribution, we systematically evaluate the $\mathcal{O}(\partial)$ contributions from both Poisson bracket and intrinsic gradient correction to the self-energies. Surprisingly, the thermal vorticity contributions from these two sources cancel exactly, resulting in vanishing vorticity-induced spin alignment at the first order.

This paper is organized as follows: In Sec.~\ref{sec:Green's-function-and}, we present the two-point functions defined on a closed time path contour and derive the leading- and next-to-leading-order DSEs in phase space through the Wigner transformation.
In Sec.~\ref{sec:self-energy}, we present the explicit form of self-energies at leading and next-to-leading order. In Sec.~\ref{sec:solution-of-DSE}, we give the formal solutions to DSEs and extract the spin alignment. Numerical results for the $\rho$ meson are presented in Sec.~\ref{sec:numerical-calculation}.

In this paper, we adopt the following notations: $g^{\mu\nu}=\text{diag}\left(1,-1,-1,-1\right)$ where $\mu,\nu=0,1,2,3$, $x^{\mu}=\left(x^{0},\mathbf{x}\right)=\left(x_{0},\mathbf{x}\right)$, $x\cdot y=x^{\mu}y^{\nu}g_{\mu\nu}$, $x^{(\mu}y^{\nu)}=\frac{1}{2}\left(x^{\mu}y^{\nu}+x^{\nu}y^{\mu}\right)$, $x^{[\mu}y^{\nu]}=\frac{1}{2}\left(x^{\mu}y^{\nu}-x^{\nu}y^{\mu}\right)$, $\hbar=k_{B}=c=1$. The four-momentum $p^{\mu}$ is not necessarily on-shell unless we add an index 'on'. The thermal velocity is given by $\beta_\mu=\beta_0u_\mu$ with $\beta_0$ and $u_\mu$ denote the inverse temperature in local rest frame and velocity, respectively. Thermal vorticity $\omega_{\mu\nu}$ and thermal shear $\xi_{\mu\nu}$ are defined as $\omega_{\mu\nu}=\partial_{[\mu}\beta_{\nu]}$, $\xi_{\mu\nu}=\partial_{(\mu}\beta_{\nu)}$. The projector operators are defined as $\Delta^{\mu\nu}_{p}=g^{\mu\nu}-\frac{p^\mu p^\nu}{p^2}$, $\Delta_{L}^{\mu\nu}=\frac{\Delta_{p}^{\mu\rho}\Delta_{p}^{\nu\sigma}u_{\rho}u_{\sigma}}{\Delta_{p}^{\rho\sigma}u_{\rho}u_{\sigma}}$, $\Delta_{T}^{\mu\nu}=\Delta_{p}^{\mu\nu}-\Delta_{L}^{\mu\nu}$. The quantities with and without tilde refer to those in the local rest frame and lab frame respectively. The summation of repeated index is implied if not stated explicitly.

\section{Green's functions and Dyson-Schwinger equations on closed-time path
\label{sec:Green's-function-and}}

In this section, we will briefly introduce the Green's function and Dyson-Schwinger equation (DSE) on the closed-time path contour (CTP)~\citep{Chou:1984es,Blaizot:2001nr,Wang:2001dm,Berges:2004yj,Cassing:2008nn,Gao:2012ix,Chen:2012ca,Crossley:2015evo,Sheng:2021kfc,Weickgenannt:2019dks,Weickgenannt:2020aaf,Weickgenannt:2021cuo}. 
Depending on which branch the field lies on, two-point Green's function $G_{CTP}$ has four components~\citep{Yin:2024dnu}, i.e. $G^{i}_{\mu\nu}$ with $i=F,\bar{F},<,>$.
Owing to the constraint $G_{\mu\nu}^{F}+G_{\mu\nu}^{\overline{F}}=G_{\mu\nu}^{<}+G_{\mu\nu}^{>}$, only three of them are independent. In this paper, we select three independent components as $G^{<}$, $G^{R}=G^F-G^<$ and $G^{A}=G^F-G^>$, where the latter two are retarded and advanced Green's functions which are known as physical representations~\citep{kadanoff1962quantum,Chou:1984es,zubarev1996statistical,zubarev1997statistical}.

The DSEs for these three independent components on CTP read 
\begin{align}
G_{A/R}^{\mu\nu}\left(x_{1},x_{2}\right)= &\; G_{A/R}^{\mu\nu(0)}\left(x_{1},x_{2}\right)+\int dx_{3}dx_{4} \left( G_{A/R}^{(0)}\star\Sigma_{A/R}\star G_{A/R}\right)^{\mu\nu} \left(x_{1},x_{2}\right)\;,\nonumber \\
G_{<}^{\mu\nu}\left(x_{1},x_{2}\right)= &\; G_{<}^{\mu\nu(0)}\left(x_{1},x_{2}\right)+\int dx_{3}dx_{4}\left(G_{R}^{(0)}\star\Sigma_{R}\star G_{<}\right)^{\mu\nu}\left(x_{1},x_{2}\right)\nonumber \\
 & +\int dx_{3}dx_{4}\left(G_{R}^{(0)}\star\Sigma_{<}\star G_{A}\right)^{\mu\nu}\left(x_{1},x_{2}\right)\nonumber \\
 & +\int dx_{3}dx_{4}\left(G_{<}^{(0)}\star\Sigma_{A}\star G_{A}\right)^{\mu\nu}\left(x_{1},x_{2}\right)\;,\label{eq:DSE_G_<}
\end{align}
where we used the short-hand notation  $(A\star B\star C)^{\mu\nu}\left(x_{1},x_{2}\right)=  A_{\;\;\rho}^{\mu}\left(x_{1},x_{3}\right)B_{\;\;\sigma}^{\rho}\left(x_{3},x_{4}\right)C^{\sigma\nu}\left(x_{4},x_{2}\right)$.

Applying the Wigner transformation $G(X,p)=\int d^4x e^{ipy}G(X+y/2,X-y/2)$ with $X=(x_1+x_2)/2$ and $y=x_1-x_2$, we can rewrite the DSE in phase space and expand it in powers of space-time gradients. This expansion is equivalent to the $\hbar$ expansion, which is widely used in spin kinetic theory~\citep{Gao:2012ix,Hidaka:2016yjf,Sheng:2021kfc,Fang:2022ttm,Dong:2022yzt}. 
The DSEs at the leading and next-to-leading order read
\begin{eqnarray}
G_{A/R}^{\mu\nu(\text{L/NL})}&= & \mathcal{F}_{A/R}^{\mu\nu(\mathrm{L/NL})}+G_{A/R}^{\mu\rho(0,\text{L})}\Sigma_{\rho\sigma}^{A/R(\text{L})}G_{A/R}^{\sigma\nu(\text{L/NL})}\;,\label{eq:G_A}\\
G_{<}^{\mu\nu(\text{L/NL})}&= & \mathcal{F}_{<}^{\mu\nu(\text{L/NL})}+G_{R}^{\mu\rho(0,\mathrm{L})}\Sigma_{\rho\sigma}^{R(\text{L})}G_{<}^{\sigma\nu(\text{L/NL})}\;,\label{eq:G_<}
\end{eqnarray}
where the explicit dependence of $\mathcal{F}$, $G$ and $\Sigma$ on $(X,p)$ are suppressed for simplicity. 
The superscripts 'L' and 'NL' stand for the leading and next-to-leading order in space-time gradients, respectively. 
The explicit form of $\mathcal{F}$ is presented in the Appendix~\ref{appendix:DSE}. 
As shown in Eq.~(\ref{eq:F}), the function $\mathcal{F}_{A/R}$ only involves the corresponding advanced/retarded component, which allows the advanced and retarded parts of Eq.~(\ref{eq:G_A}) to be solved separately. 
Its solution in leading order is related to the spectral function of the vector meson~\citep{Dong:2023cng}, and the NLO solution represents the $\mathcal{O}(\partial)$ correction to the spectral function. 

As presented explicitly in Eq.~(\ref{eq:F}), the $\mathcal{F}_<^{\text{NL}}$ contains contributions from Poisson brackets, gradient corrections to the self-energies, and NLO retarded/advanced Green's functions. The first source is well studied in kinetic theory, while the latter two are rarely discussed.

\section{Self-energies \label{sec:self-energy}}

\begin{figure}
\begin{centering}
\includegraphics{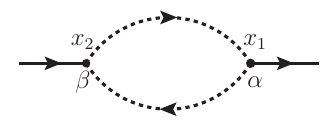}\includegraphics{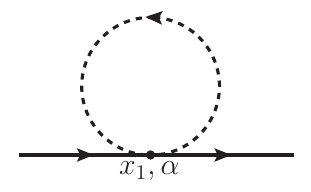}
\par\end{centering}
\caption{Self-energies of the vector meson at one-loop level. The dashed lines are propagators of pions, while the solid external lines are $\rho$ mesons. \label{fig:feynmann-diagrams}}
\end{figure}

To solve DSEs given in Eq.~(\ref{eq:G_A}) and Eq.~(\ref{eq:G_<}), the explicit form of the self-energy is required. For the purpose of illustration, we take the interaction between $\rho$ mesons and pions as an example. The corresponding Lagrangian derived from the chiral perturbation theory~\citep{Gale:1990pn,Klingl:1996by,Scherer:2002tk} reads 
\begin{equation}
\mathcal{L} = \mathcal{L}_{\mathrm{kin}}+ig_{\rho\pi}\left(\partial_{\mu}\pi^{+}\pi^{-}-\pi^{+}\partial_{\mu}\pi^{-}\right)\rho^{\mu}+g_{\rho\pi}^{2}\pi^{+}\pi^{-}\rho_{\mu}\rho^{\mu}\;,
\label{eq:Chpt-lagrangian} 
\end{equation}
where $\mathcal{L}_{\mathrm{kin}}$ denotes the kinetic part of the mesons, and the coupling constant is extracted from the decay width of the $\rho$ meson. 
For illustration, we evaluate the self-energies at the one-loop level. The corresponding Feynman diagrams generated by the $\rho\pi\pi$ and $\rho\rho\pi\pi$ couplings are shown in Fig.~\ref{fig:feynmann-diagrams}.

The self-energies on the CTP contour can then be written as 
\begin{align}
\Sigma_{\alpha\beta}^{<}\left(x_{1},x_{2}\right)= & g_{\rho\pi}^{2}\left(\partial_{\alpha}^{1}S_{12}^{<}\partial_{\beta}^{2}S_{21}^{>}+\partial_{\alpha}^{1}S_{21}^{>}\partial_{\beta}^{2}S_{12}^{<}\right)\nonumber \\
 & -g_{\rho\pi}^{2}\left[\left(\partial_{\alpha}^{1}\partial_{\beta}^{2}S_{12}^{<}\right)S_{21}^{>}+\left(\partial_{\alpha}^{1}\partial_{\beta}^{2}S_{21}^{>}\right)S_{12}^{<}\right]\;,\nonumber\\
\Sigma_{\alpha\beta}^{F}\left(x_{1},x_{2}\right)= & g_{\rho\pi}^{2}\left(\partial_{\alpha}^{1}S_{12}^{F}\partial_{\beta}^{2}S_{21}^{F}+\partial_{\alpha}^{1}S_{21}^{F}\partial_{\beta}^{2}S_{12}^{F}\right)\nonumber \\
 & -g_{\rho\pi}^{2}\left[\left(\partial_{\alpha}^{1}\partial_{\beta}^{2}S_{12}^{F}\right)S_{21}^{F}+\left(\partial_{\alpha}^{1}\partial_{\beta}^{2}S_{21}^{F}\right)S_{12}^{F}\right]\nonumber \\
 & +2ig_{\rho\pi}^{2}g_{\alpha\beta}S_{12}^{F}\delta\left(x_{1}-x_{2}\right)+\Sigma_{\alpha\beta}^{\text{ren}}\;,\label{eq:sigma_F}
\end{align}
where we used the notations $S_{12}=S\left(x_{1},x_{2}\right)$ and $\partial^{1/2}=\partial/\partial x_{1/2}$. 
The renormalization term $\Sigma_{\alpha\beta}^{\text{ren}}$ is introduced to cancel the ultraviolet divergence with the renormalization condition $\text{Im}\Sigma^{F}\left(p^{2}=m_{\rho}^{2}\right)=0$~\citep{Gale:1990pn}. 
Applying the Wigner transformation to Eq.~(\ref{eq:sigma_F}), the self-energies can be expanded in power of space-time gradients and the leading-order results~\citep{Yin:2024dnu,Yin:2025gvl} are expressed in a compact form
\begin{equation}
\Sigma_{\mu\nu}^{R(\text{L})}  =-\underset{a=L,T}{\sum}\Delta_{\mu\nu}^{a}\Sigma_{a},\;\;\; 
\Sigma_{\mu\nu}^{<(\text{L})}  =-\underset{a=L,T}{\sum}\Delta_{\mu\nu}^{a}\Sigma_{a}^{<},
\label{eq:Sigma_R/<_L}
\end{equation}
where the explicit derivation of transverse/longitudinal self-energy $\Sigma_{L/T}$ is presented in Appendix ~\ref{appendix:calculation-of-self}. 
The projector operators $\Delta^{\mu\nu}_{L}$ and $\Delta^{\mu\nu}_{T}$ are orthogonal to $p^\mu$ and project a vector onto the direction parallel and transverse to $\Delta^{\mu\nu}_{p}u_\nu$, respectively. 
In contrast to the vacuum case, in which self-energy is proportional to $\Delta^{\mu\nu}$, the flow velocity $u^\mu$ in a thermal medium introduces an additional preferred direction. This allows the self-energies to be decomposed into transverse and longitudinal parts. 
A more general decomposition of $\Sigma^{\text{(L)}}_{\mu\nu}$ should include the term proportional to $p^{\mu/\nu}$. However, in the case of $\rho-\pi$ system, these terms vanish due to the Ward identity.
The corresponding advanced self energy can be then obtained via $\Sigma_{A}^{\mu\nu(\text{L})}=-\Sigma_{R}^{\mu\nu(\text{L})*}$ because of the symmetry between the retarded and advanced components.

At the next-to-leading order in gradient expansion, the self-energies contain two kinds of corrections. 
The first arises from the next-to-leading-order corrections to pion distribution, while the second from the explicit gradient terms generated by Wigner transformation of self-energies themselves. We refer to the latter one as the intrinsic gradient correction to self-energies. The former depends on the specific non-equilibrium form of the pion distributions which is still an open question~\citep{Fang:2025pzy}. Therefore, in this paper, we apply the locally equilibrated pion distribution and mainly focus on the second source. 
Performing the Wigner transformation to the self-energy, the NLO corrections to self-energy yield
\begin{align}
\Sigma_{\alpha\beta}^{<(\text{NL})}= & -2\partial_{[\alpha}^{X}\beta_{\beta]}H_{1}-2u_{[\beta}\partial_{\alpha]}^{X}\beta_{\rho}p^{\rho}H_{2}-2u_{[\beta}\partial_{\alpha]}^{X}\beta_{\rho}u^{\rho}H_{3}\nonumber \\
 & +(p_{\alpha}H_{4,\beta} -p_{\beta}H_{4,\alpha})\;,\nonumber \\
\Sigma_{\alpha\beta}^{R(\text{NL})}= & -2\partial_{[\alpha}^{X}\beta_{\beta]}F_{1}-2u_{[\beta}\partial_{\alpha]}^{X}\beta_{\rho}p^{\rho}F_{2}-2u_{[\beta}\partial_{\alpha]}^{X}\beta_{\rho}u^{\rho}F_{3}\nonumber \\
 & +(p_{\alpha}F_{4,\beta} -p_{\beta}F_{4,\alpha})\;,\label{eq:sigma_NL}
\end{align}
where $H_{i}$ and $F_{i}$ are functions of $X^{\mu}$ and $p^{\mu}$ and are given in Appendix~\ref{appendix:self-energy-correction} ($H_{4}$ and $F_{4}$ are excluded because they do not contribute to the final result). 
The NLO corrections to $\Sigma_{\alpha\beta}^{<}$ and $\Sigma_{\alpha\beta}^{R}$ are antisymmetric under the interchange of $\alpha$ and $\beta$. Therefore, the advanced self-energy can be obtained from $\Sigma_{\mu\nu}^{A(\text{NL})}=-\Sigma_{\nu\mu}^{R(\text{NL})*}=\Sigma_{\mu\nu}^{R(\text{NL})*}$.

Equation (\ref{eq:sigma_NL}) explicitly shows that the derivative couplings between $\rho$ mesons and pions can generate non-vanishing NLO corrections to self-energies under the Wigner transformation. 
Such gradient corrections also exist in other theories like QED/QCD when taking the higher-loop self energy diagrams into consideration.

\section{Solutions to DSE and spin alignments \label{sec:solution-of-DSE}}

In this section, we will solve the leading and next-to-leading order DSE with self-energies obtained in the Sec.~\ref{sec:self-energy}, and then determine the spin alignment of $\rho^{0}$ mesons analytically.

\subsection{Leading-order solution\label{subsec:Leading-order-solution}}
The method for solving the DSE is presented in Appendix~\ref{appendix:DSE}. The leading-order solutions to Green's functions are 
\begin{align}
G^{R(\text{L})}_{\mu\nu} & =-i\underset{a=T,L}{\sum}\Delta^{a}_{\mu\nu}\rho_{a}+i\frac{p_{\mu}p_{\nu}}{p^{2}m_{V}^{2}}\;,\label{eq:G_L_R_solution}\\
G_{\mu\nu}^{<(\text{L})} & =\underset{a=T,L}{\sum}\Delta_{\mu\nu}^{a}\rho_{a}\rho_{a}^{*}\Sigma^{<}_{a}\;.
\label{eq:G_L_<_solution}
\end{align}
where the special functions $\rho_{L/T}$ are defined by 
\begin{equation}
\rho_{L/T}=\frac{1}{p^{2}-m_{V}^{2}-i\Sigma_{L/T}}\;.
\end{equation}
Here, the transverse and longitudinal parts of self-energies $\Sigma_{L/T}$ are given in Eq.~(\ref{eq:sigma_L/T}). The imaginary part of $\rho_{T/L}$ denotes the spectral function of vector mesons.
Due to the symmetry between retarded and advanced Green's functions, $G^{A(\text{L})}_{\mu\nu}$ can be obtained by taking the complex conjugate of $G^{R(\text{L})}_{\mu\nu}$, i.e. $G_{\mu\nu}^{A(\text{L})}=-G_{\mu\nu}^{R(\text{L})*}$. 
The minus sign comes from the imaginary unit $i$ in the numerator of Eq.~(\ref{eq:G_A_(0)}). 
Since pions are spinless and their distributions are assumed to be locally equilibrated, the leading-order self-energies satisfy the Kubo-Martin-Schwinger (KMS) relation $\Sigma_{\mu\nu}^{>(\text{L})}=e^{\beta p\cdot u}\Sigma_{\mu\nu}^{<(\text{L})}$. This allows us to rewrite the self-energy as $\Sigma^{<}_{a} =2n_{B}\mathrm{Re}\left(i/\rho_{a}\right)$, where $n_{B}=1/\left(e^{\beta p\cdot u}-1\right)$ denotes the Bose-Einstein distribution. 
Substituting this relation into Eq.~(\ref{eq:G_L_<_solution}), we obtain 
\begin{align}
G_{\mu\nu}^{<(\text{L})} & =2n_{B}\underset{a=T,L}{\sum}\Delta_{\mu\nu}^{a}\mathrm{Re}\left(i\rho_{a}^{*}\right),
\label{eq:G_L_<_solution2}
\end{align}
which is the well-known fluctuation-dissipation theorem~\citep{fetter2003quantum,zubarev1997statistical}. 
It should be emphasized that Eq.~(\ref{eq:G_L_<_solution2}) relies on the local-equilibrium assumption. For more general distribution functions, the $G_{\mu\nu}^{<}$ should instead be evaluated using Eq.~(\ref{eq:G_L_<_solution})~\citep{Fang:2026ztm}.

\subsection{Next-to-leading order solutions}

Following the similar method with the leading order, NLO Dyson equations can be solved, and the results are  
\begin{align}
G_{\mu\nu}^{R(\text{NL})}= & G_{\mu\rho}^{R(\text{L})}\Sigma_{R}^{\rho\sigma(\text{NL})}G_{\sigma\nu}^{R(\text{L})}-i\frac{p_{\alpha}}{\left(p^{2}-m_{V}^{2}\right)}G_{\mu\rho}^{R(\text{L})}\partial_{X}^{\alpha}\left[\Sigma_{R}^{\rho\sigma(\text{L})}G_{\sigma\nu}^{R(\text{L})}\right]\nonumber \\
 & -\frac{1}{2}\frac{p_{\mu}}{m_{V}^{2}\left(p^{2}-m_{V}^{2}\right)}\partial_{\rho}^{X}\left[\Sigma_{R}^{\rho\sigma(\text{L})}G_{\sigma\nu}^{R(\text{L})}\right]-\frac{i}{2}G_{\mu\rho}^{R(\text{L})}\left[\Sigma_{R}^{\rho\sigma(\text{L})},G_{\sigma\nu}^{R(\text{L})}\right]_{P.B.}\;,\label{eq:G_NL_R_solution}\\
G_{\mu\nu}^{<(\text{NL})}= & G_{\mu\rho}^{R(\text{L})}\Sigma_{<}^{\rho\sigma(\text{NL})}G_{\sigma\nu}^{A(\text{L})}+G_{\mu\rho}^{R(\text{L})}\Sigma_{R}^{\rho\sigma(\text{NL})}G_{\sigma\beta}^{R(\text{L})}\Sigma_{<}^{\beta\alpha(\text{L})}G_{\alpha\nu}^{A(\text{L})}\nonumber \\
 & +G_{\mu\rho}^{R(\text{L})}\Sigma_{<}^{\rho\sigma(\text{L})}G_{\sigma\beta}^{A(\text{L})}\Sigma_{A}^{\beta\alpha(\text{NL})}G_{\alpha\nu}^{A(\text{L})}\nonumber \\
 & +\frac{i}{2}G_{\mu\rho}^{R(\text{L})}\left[G_{R}^{-1,\rho\sigma(\text{L})},G_{\sigma\nu}^{<(\text{L})}\right]_{P.B.}+\frac{i}{2}\left[G_{\mu\alpha}^{<(\text{L})},G_{A}^{-1,\alpha\sigma(\text{L})}\right]_{P.B.}G_{\sigma\nu}^{A(\text{L})}\nonumber \\
 & +\frac{i}{2}G_{\mu\rho}^{R(\text{L})}G_{\beta\alpha}^{<(\text{L})}\left[G_{R}^{-1,\rho\beta(\text{L})},G_{A}^{-1,\alpha\sigma(\text{L})}\right]_{P.B.}G_{\sigma\nu}^{A(\text{L})}\nonumber \\
 & +p_{\mu/\nu}\text{ proportional terms}\;.\label{eq:G_NL_<_solution}
\end{align}
Equation (\ref{eq:G_NL_<_solution}) provides a formal NLO solution to the DSE for a general class of interactions whose leading-order self-energies can be decomposed into transverse and longitudinal part like Eq.~(\ref{eq:Sigma_R/<_L}). It explicitly shows that, in addition to Poisson bracket terms, the NLO corrections to self-energies can contribute to the spin alignment.

\subsection{Spin density matrices and spin alignments}
As shown in Eq.~(\ref{eq:G_<_(0)}), the matrix-valued spin dependent distribution (MVSD) of a free vector meson can be extracted by contracting $\epsilon^{\mu*}\left(\lambda_{1},p\right)\epsilon^{\nu}\left(\lambda_{2},p\right)$ to $G^{<}_{\mu\nu}$. We assume that this relation could be generalized to the interacting system, allowing us to calculate the MVSD as  
\begin{align}
f_{\lambda_{1}\lambda_{2}}^{V}\left(X,\boldsymbol{p}\right) & =\int_{0}^{\infty}\frac{dp^{0}}{2\pi}2p^{0}\epsilon^{\mu *}\left(\lambda_{1},p\right)G^{<}_{\mu\nu}\left(X,p\right)\epsilon^{\nu}\left(\lambda_{2},p\right)\;,\label{eq:f_0}
\end{align}
where the integration range of $p^{0}$ is restricted to $[0,\infty)$ to retain only the positive-energy contribution. 
It should be emphasized that, although this formula is widely adopted, this extension to interacting systems remains an ansatz and the exact correspondence between the $G^<$ and $f_{\lambda_1 \lambda_2}$ in an interacting system has not been fully established. 
Inserting Eqs. (\ref{eq:G_L_<_solution}) and (\ref{eq:G_NL_<_solution}) into Eq.~(\ref{eq:f_0}), the '00' component and trace of $f_{\lambda_{1}\lambda_{2}}$, i.e. $\mathrm{Tr} (f^V)$, can be then evaluated as 
\begin{align}
f_{00}^{V} & =\;\mathcal{T}^{0}-\partial^{\rho}\beta_{0}\mathcal{T}_{\rho}^{1}-\xi^{\rho\sigma}\mathcal{T}_{(\rho\sigma)}^{2}-\omega^{\rho\sigma}\mathcal{T}_{[\rho\sigma]}^{3}\;,\label{eq:f_00}\\
\mathrm{Tr}\left(f^V\right) & =\;\mathcal{N}^{0}+\partial^{\rho}\beta_{0}\mathcal{N}_{\rho}^{1}+\xi^{\rho\sigma}\mathcal{N}_{(\rho\sigma)}^{2}\;,
\label{eq:trace_f}
\end{align}
where the explicit form of the functions $\mathcal{T}$ and $\mathcal{N}$ are given in Eq.~(\ref{eq:T_N}). 
The $\mathcal{T}^{0}$ and $\mathcal{N}^{0}$ characterize the leading-order distributions, which can generate non-vanishing spin alignment due to the medium induced spectral difference. A similar mechanism also exists in quark-meson systems~\citep{Dong:2023cng}. 
Equations~(\ref{eq:f_00}) and~(\ref{eq:trace_f}) show that the thermal shear $\xi^{\rho\sigma}$, thermal vorticity $\omega^{\rho\sigma}$ and temperature gradient $\partial^{\rho}\beta_{0}$ appear explicitly in $f^{\text{NLO}}_{00}$, while no thermal vorticity term appears in the spin-summed particle distribution. 
The explicit forms of $\mathcal{N}_{\rho}^{1}$ and $\mathcal{N}_{(\rho\sigma)}^{2}$ are independent of $h_{1}\cdots h_{6}$ in Table~\ref{tab:h_i}, indicating that the NLO self-energies considered here do not modify the spin-summed particle distribution.

Taking the ratio of $f_{00}^{V}$ to $\mathrm{Tr}(f^V)$ and absorbing $\mathcal{T}^{1}$ into $\mathcal{T}^{2,3}$ with the relation $\partial^{\rho}\beta_{0}=\left(\omega^{\rho\sigma}+\xi^{\rho\sigma}\right)u_{\sigma}$, we obtain 
\begin{equation}
\rho_{00}\left(X,\mathbf{p}\right)-\frac{1}{3}\approx \delta\rho_{00}^{(0)}-\xi^{\rho\sigma}X_{(\rho\sigma)}-\omega^{\rho\sigma}Y_{[\rho\sigma]}\;.\label{eq:rho_00-1/3}
\end{equation}
where the coefficients are defined as
\begin{align}
\delta\rho_{00}^{(0)}  =&\frac{\mathcal{T}^{0}-\mathcal{N}^{0}/3}{\mathcal{N}^{0}}\;,Y_{\rho\sigma}=0\;, \nonumber\\
X_{\rho\sigma}  =&\frac{\mathcal{T}_{(\rho\sigma)}^{2}+\mathcal{T}^{1}_{(\rho}u_{\sigma)}}{\mathcal{N}^{0}}+\frac{\mathcal{N}_{(\rho\sigma)}^{2}+\mathcal{N}_{(\rho}^1 u_{\sigma)}}{3\mathcal{N}^0}\nonumber \\ &+\delta\rho_{00}^{(0)}\frac{\mathcal{N}_{(\rho\sigma)}^{2}+\mathcal{N}^{1}_{(\rho}u_{\sigma)}}{\mathcal{N}^0}\;.
\label{eq:XYZ}
\end{align}
Here, we have included the NLO contributions in the denominator using $1/(x+\delta x)\approx1/x-\delta x/x^{2}$. 
Substituting Eq.~(\ref{eq:T_N}) into Eq.~(\ref{eq:XYZ}), one can verify that the NLO spin alignment is related with the difference between 'T' and 'L' components. 
This difference, as shown in Appendix~\ref{appendix:calculation-of-self} and Appendix~\ref{appendix:self-energy-correction}, originates from the interaction with medium. In the absence of such a difference, the NLO spin alignment in the present framework vanishes. 

The structure of NLO spin alignment in this paper differs from that obtained in previous calculation via linear response theory~\citep{Li:2022vmb,Dong:2023cng,Yin:2025gvl}. In those calculations, the thermal shear contribution relies on individual $\text{Im}\left(\rho_{L/T}\right)$, while in this paper, it is related to their difference.
This difference can be understood as the different sources of the NLO self-energy corrections. In our previous calculation, thermal shear was introduced as a perturbation to the density operator, which can be associated with introducing the NLO pion distribution correction in calculating the self-energy correction.
In contrast, the present calculation focuses on the corrections arising from the intrinsic gradient correction to the self-energies. A complete evaluation of the first-order gradient corrections would require both sources and will be discussed in a forthcoming paper.

The coefficient $Y_{\rho\sigma}$ contains the contributions from Poisson bracket and the NLO self-energy. Although the Poisson bracket terms and the NLO self-energy corrections individually induce non-vanishing thermal vorticity contributions, they will cancel when taking the summation. Consequently, thermal vorticity does not induce spin alignment at first order. 
This cancellation is consistent with the symmetry analysis based on parity and time reversal~\citep{Yang:2024fkn}, indicating that the NLO self-energy corrections are important in preserving the underlying physical symmetries.

\section{Numerical calculation\label{sec:numerical-calculation}}

\begin{figure}
\begin{centering}
\includegraphics[scale=0.6]{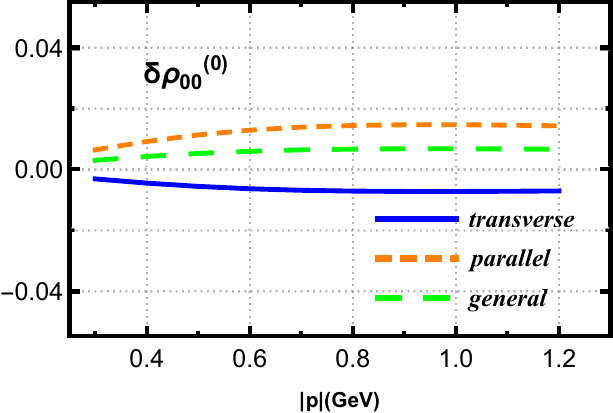}
\par\end{centering}
\caption{Leading order contribution to the spin alignment. \label{fig:rho00^0}}
\end{figure}

\begin{figure}
\begin{centering}
\includegraphics[scale=0.75]{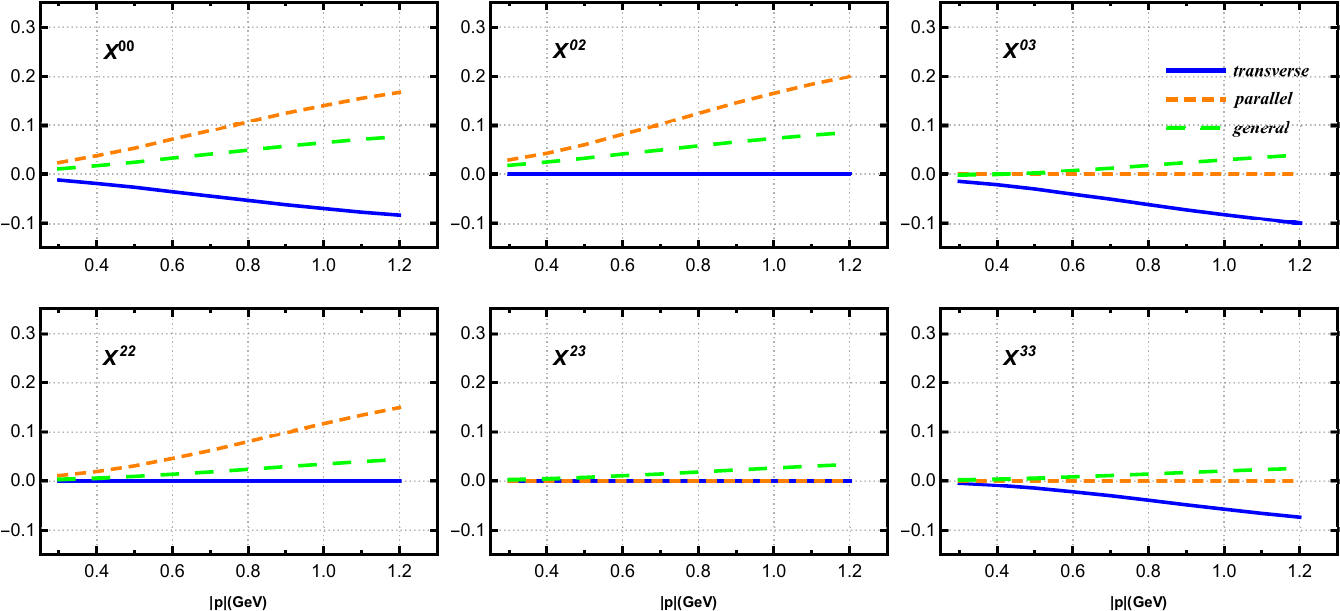}
\par\end{centering}
\caption{The numerical result of $X^{(\rho\sigma)}$. The other components are all zero.\label{fig:X}}
\end{figure}

In this section, we perform a numerical calculation to the coefficients in the local rest frame. 
As shown in Eq.~(\ref{eq:rho_00-1/3}), the integrand functions in coefficients $X^{\rho\sigma}$ are Lorentz covariant. Therefore, their value in the lab frame can be obtained through a Lorentz transformation with the boost matrix defined in Eq.~(\ref{eq:Lambda}). 
The mass of $\rho^0$ meson and pion are set to $m_{\rho}=\text{0.77 GeV}$ and $m_{\pi}=\text{0.14 GeV}$, respectively. The temperature is chosen to be $T_{f}=$0.12 GeV, which lies between the chemical and the kinetic freeze-out temperatures. 
The coupling constant $g_{\rho\pi}=6.067$ is extracted from the vacuum decay width of $\rho^0$ meson. The calculation of coefficients is performed in local rest frame, in which the projector operator contracted with polarization vector can be rewritten as 
\begin{align}
\Delta_{\mu}^{T,\rho}u_{\nu}\epsilon_{0}^{\mu}\epsilon_{0}^{\nu} & =\left(0,-\frac{\mathbf{p}_{x}\mathbf{p}_{y}^{2}}{|\mathbf{p}|^{2}\sqrt{p^{2}}},\frac{\mathbf{p}_{y}\left(\mathbf{p}_{x}^{2}+\mathbf{p}_{z}^{2}\right)}{|\mathbf{p}|^{2}\sqrt{p^{2}}},-\frac{\mathbf{p}_{z}\mathbf{p}_{y}^{2}}{|\mathbf{p}|^{2}\sqrt{p^{2}}}\right)^{\rho},\label{eq:operator-1}\\
\Delta_{\mu\nu}^{L}\epsilon_{0}^{\mu}\epsilon_{0}^{\nu}+\frac{1}{3} & =\frac{1}{3}-\frac{\mathbf{p}_{y}^{2}}{|\mathbf{p}|^{2}}\;.\label{eq:operators}
\end{align}
Then, one can calculate the leading order of spin alignment $\delta\rho_{00}^{(0)}$ and the associated coefficients $X$ numerically by Eq.~(\ref{eq:XYZ}). The results are presented in Fig.~\ref{fig:rho00^0} and Fig.~\ref{fig:X}. The labels "parallel" and "transverse" refer to cases in which three-momentum $\mathbf{p}$ is aligned parallel or transverse to the spin quantization axis $\boldsymbol{\epsilon}_{0}$ (see Eq.~(\ref{eq:epsilon_rest})). The "general" case corresponds to $p^{\mu}=\left(p^{0},0,0.8|\mathbf{p}|,0.6|\mathbf{p}|\right)$.
As shown explicitly in Eq.~(\ref{eq:T_N}), the coefficients are momentum dependent, which may enhance the magnitude of gradient corrections at large momenta. Consequently, the gradient expansion might be less reliable in the high-momentum region. Therefore, we restrict our numerical calculation to the region $|\mathbf{p}|<1.2\text{ GeV}$. 

As presented in Fig.~\ref{fig:rho00^0}, the leading-order spin alignment is of the order $\mathcal{O}\left(10^{-2}\right)$ and the parallel case has the opposite sign to the transverse one. 
Equation~(\ref{eq:operators}) further shows that the leading-order contribution vanishes under the integration over the whole momentum space. 
Therefore, the leading-order spectral difference does not contribute to the momentum-integrated spin alignment. 

As shown in Fig.~\ref{fig:X}, the coefficients in the parallel case differ from their transverse counterparts by a factor of -2. This implies that a non-vanishing momentum-integrated spin alignment requires an anisotropic thermal shear tensor.

Given the magnitude of these coefficients, the spin alignment of $\rho^{0}$ meson at NLO is expected to be of order $10^{-2}\sim10^{-3}$ when the magnitude of thermal shear, as estimated from blast-wave model~\citep{Bondorf:1978kz,Siemens:1978pb,Schnedermann:1993ws,Retiere:2003kf,Arslan:2024dwi}, reaches the order $10^{-1}\sim10^{-2}$. A quantitative prediction based on the thermal shear extracted from hydrodynamic simulation will be presented in a forthcoming paper.

\section{discussion and conclusion}

In this work, we have developed a formalism for studying the spin alignment of vector mesons based on the direct solution of the Dyson-Schwinger equations on the CTP. Through the Wigner transformation, these equations are systematically expanded into leading and next-to-leading orders in space-time gradient. The leading-order equations reproduce the transverse and longitudinal components of the vector meson propagators whose imaginary parts are related to the spectral functions. The next-to-leading order equations involve both Poisson brackets and gradient correction to self-energies. Applying this formalism to the $\rho^0$ meson in a pion gas, we find that the thermal shear $\xi^{\mu\nu}$ can induce non-vanishing spin alignment at the first order. The corresponding coefficients are closely related to the difference between transverse and longitudinal components. 

It is worth noting that, although the thermal vorticity contributions from the Poisson bracket terms and the self-energy corrections are individually non-vanishing, their summation cancel exactly. This result is in agreement with the parity and time reversal analysis \citep{Yang:2024fkn}. It indicates that the self-energy corrections can not be neglected and play an important role in keeping the underlying symmetry.

In contrast to the widely adopted quasi-particle approximation which eliminates independent closed-time Green's functions from ($G^{R}$, $G^A$, $G^<$) to ($G^<$, $G^>$), the present formalism retains the full $G^{R}$, $G^A$ and $G^<$. It allows us to investigate the off-shell effect and gradient corrections to spin-related observables systematically.
Although we restrict our calculation on the $\rho-\pi$ interaction with locally equilibrated pions here, the formalism can be extended to other vector mesons, interactions and distributions. 

The present method can also be extended to spin-1/2 particles whose Green's functions and Dyson-Schwinger equations are expressed by $4\times4$ matrix in Dirac space. 
Although the calculation is more challenging, this formalism provides a systematic framework for investigating the spin polarization of fermions beyond the quasi-particle approximation.

\begin{acknowledgments}
W.-B. Dong and Y.-L. Yin thank the local
support by the Institute for Theoretical Physics at Goethe University
Frankfurt where part of the work was done and support by China Scholarship
Council. W.-B. Dong, Y.-L. Yin and Q. Wang are supported in part by
National Natural Science Foundation of China under the grant No. 12135011.
\end{acknowledgments}

\appendix
\section{Dyson-Schwinger equations and their formal solutions\label{appendix:DSE}}
Applying the Wigner transformation to Eq.~(\ref{eq:DSE_G_<}), we obtain the special function $\mathcal{F}$ at its leading and next-to-leading order, yielding
\begin{align}
\mathcal{F}^{\mu\nu}_{A/R(\text{L})}=&G_{A/R}^{\mu\nu(0,\mathrm{L})}\;,\nonumber\\
\mathcal{F}^{\mu\nu}_{<(\text{L})}=&G_{<}^{\mu\nu(0,\mathrm{L})}+G_{R}^{\mu\rho(0,\text{L})}\Sigma_{\rho\sigma}^{<(\text{L})}G_{A}^{\sigma\nu(\text{L})}+G_{<}^{\mu\rho(0,\text{L})}\Sigma_{\rho\sigma}^{A(\text{L})}G_{A}^{\sigma\nu(\text{L})}\;, \nonumber\\   \mathcal{F}^{A/R(\text{NL})}_{\mu\nu}=&G_{\mu\rho}^{A/R(0,\text{L})}\Sigma_{A/R}^{\rho\sigma(\text{NL})}G_{\sigma\nu}^{A/R(\text{L})}\nonumber \\
 & -\frac{i}{2}\left[G_{\mu\rho}^{A/R(0,\text{L})},\Sigma_{A/R}^{\rho\sigma(\text{L})}G_{\sigma\nu}^{A/R(\text{L})}\right]_{PB}\nonumber \\
 & -\frac{i}{2}G_{\mu\rho}^{A/R(0,\text{L})}\left[\Sigma_{A/R}^{\rho\sigma(\text{L})},G_{\sigma\nu}^{A/R(\text{L})}\right]_{PB}\;,\nonumber\\
\mathcal{F}^{<(\text{NL})}_{\mu\nu}=&G^{<(0,\text{NL})}_{\mu\nu}+G_{\mu\rho}^{R(0,\text{L})}\Sigma_{R}^{\rho\sigma(\text{NL})}G_{\sigma\nu}^{<(\text{L})}+G_{\mu\rho}^{<(0,\text{L})}\Sigma_{A}^{\rho\sigma(\text{NL})}G_{\sigma\nu}^{A(\text{L})}\nonumber \\
 & +G_{\mu\rho}^{<(0,\text{NL})}\Sigma_{A}^{\rho\sigma(\text{L})}G_{\sigma\nu}^{A(\text{L})}+G_{\mu\rho}^{R(0,\text{L})}\Sigma_{<}^{\rho\sigma(\text{NL})}G_{\sigma\nu}^{A(\text{L})}\nonumber \\
 & +G_{\mu\rho}^{<(0,\text{L})}\Sigma_{A}^{\rho\sigma(\text{L})}G_{\sigma\nu}^{A(\text{NL})}+G_{\mu\rho}^{R(0,\text{L})}\Sigma_{<}^{\rho\sigma(\text{L})}G_{\sigma\nu}^{A(\text{NL})}\nonumber \\
 & -\frac{i}{2}\left[G_{\mu\rho}^{R(0,\text{L})},\Sigma_{R}^{\rho\sigma(\text{L})}G_{\sigma\nu}^{<(\text{L})}\right]_{PB}-\frac{i}{2}\left[G_{\mu\rho}^{<(0,\text{L})},\Sigma_{A}^{\rho\sigma(\text{L})}G_{\sigma\nu}^{A(\text{L})}\right]_{PB}\nonumber \\
 & -\frac{i}{2}\left[G_{\mu\rho}^{R(0,\text{L})},\Sigma_{<}^{\rho\sigma(\text{L})}G_{\sigma\nu}^{A(\text{L})}\right]_{PB}-\frac{i}{2}G_{\mu\rho}^{R(0,\text{L})}\left[\Sigma_{R}^{\rho\sigma(\text{L})},G_{\sigma\nu}^{<(\text{L})}\right]_{PB}\nonumber \\
 & -\frac{i}{2}G_{\mu\rho}^{<(0,\text{L})}\left[\Sigma_{A}^{\rho\sigma(\text{L})},G_{\sigma\nu}^{A(\text{L})}\right]_{PB}-\frac{i}{2}G_{\mu\rho}^{R(0,\text{L})}\left[\Sigma_{<}^{\rho\sigma(\text{L})},G_{\sigma\nu}^{A(\text{L})}\right]_{PB}\;.\label{eq:F}
\end{align}
The subscript 'PB' denotes the Poisson brackets, defined as $\left[A,B\right]_{PB}=\partial_{x}^{\mu}A\partial_{\mu}^{p}B-\partial_{\mu}^{p}A\partial_{x}^{\mu}B$. 
Propagators in non-interacting systems are given by
\begin{align}
G_{R/A}^{\mu\nu(0)}\left(x,p\right)= & \left(g^{\mu\nu}-\frac{p^{\mu}p^{\nu}}{m_{V}^{2}}\right)\frac{-i}{p^{2}-m_{V}^{2}\pm i\epsilon p^{0}}\;,\label{eq:G_A_(0)}\\
G_{\mu\nu(0)}^{<}\left(x,p\right)= & 2\pi\epsilon_{\mu}\left(\lambda_{1},p\right)\epsilon_{\nu}^{*}\left(\lambda_{2},p\right)\theta\left(p^{0}\right)\delta\left(p^{2}-m_{V}^{2}\right)f_{\lambda_{1}\lambda_{2}}^{V(0)}\left(x,\boldsymbol{p}\right)\nonumber \\
 & +2\pi\epsilon_{\mu}^{*}\left(\lambda_{1},-p\right)\epsilon_{\nu}\left(\lambda_{2},-p\right)\theta\left(-p^{0}\right)\delta\left(p^{2}-m_{V}^{2}\right)\left[\delta_{\lambda_1\lambda_2}+f_{\lambda_{2}\lambda_{1}}^{V(0)}\left(x,-\boldsymbol{p}\right)\right]\;,\label{eq:G_<_(0)}
\end{align}
where the $f_{\lambda_{1}\lambda_{2}}^{V(0)}$ is the matrix-valued spin distribution (MVSD) of the vector meson in the non-interacting limit~\citep{Becattini:2013fla,Sheng:2021kfc}. According to kinetic theory, the on-shell condition of $G^{\mu\nu(0)}_<$ holds in the leading and next-to-leading order of the space-time gradient~\citep{Sheng:2021kfc,Sheng:2022ffb}. 

With the explicit form of free propagators, the leading-order DSE for retarded component can be rewritten as
\begin{align}
    G^{R(\text{L}),-1}_{\mu\nu}=&G^{R,(\text{L}),-1}_{(0),\mu\nu}-\Sigma^{R(\text{L})}_{\mu\nu}\nonumber\\
    =&i(p^2\Delta^p_{\mu\nu}-m_V^2g_{\mu\nu}-i\Delta^T_{\mu\nu}\Sigma^T-i\Delta^L_{\mu\nu}\Sigma^L)\;.\label{eq:G^R,-1}
\end{align}
The solution can be obtained by taking the inverse of Eq.~(\ref{eq:G^R,-1}) and the result reads
\begin{align}
    G^{R(\text{L})}_{\mu\nu}=-i\underset{a=L,T}{\sum}\Delta_{\mu\nu}^a\rho_a+i\frac{p_\mu p_\nu}{m_{V}^2p^2}\;.
\end{align}
With the constraint $(g-G_{R}^{\text{(0,L)}}\Sigma^{(\text{L})}_R)^{-1}_{\mu\nu}=(G^{(\text{L})}_{R}G^{(0,\text{L}),-1}_{R})_{\mu\nu}$, the DSE can be solved by 
\begin{align}
    G^{R\text{(NL)}}_{\mu\nu}=&(G_{R}^{\text{(L)}}G_{R}^{\text{(0,L)},-1}\mathcal{F}_R^{\text{(NL)}})_{\mu\nu}\;,\nonumber\\G^{<\text{(L/NL)}}_{\mu\nu}=&(G_{R}^{\text{(L)}}G_{R}^{\text{(0,L)},-1}\mathcal{F}_<^{\text{(L/NL)}})_{\mu\nu}\;.\label{eq:method_solve}
\end{align}
Together with the on-shell condition of $G_<^{(0)}$, the Eq.~(\ref{eq:method_solve}) can be simplified into the solution we presented in Sec.~\ref{sec:solution-of-DSE} 
\section{Calculation of leading-order self-energies\label{appendix:calculation-of-self} }

The fields of $\rho$ and $\pi$ mesons can be expanded in terms of creation and annihilation operators as
\begin{align}
\rho^{\mu}(x) & =\sum_{\lambda=0,\pm1}\int\frac{d^{3}\mathbf{p}}{(2\pi)^{3}}\frac{1}{2E_{p}^{\rho}}\left[\epsilon^{\mu}(\lambda,p_{\text{on}})a_{V}(\lambda,{\bf p})e^{-ip_{\mathrm{on}}\cdot x}+\epsilon^{\mu\ast}(\lambda,{\bf p}_{\text{on}})a_{V}^{\dagger}(\lambda,{\bf p})e^{ip_{\mathrm{on}}\cdot x}\right]\;,\\
\pi^{+}(x) & =\int\frac{d^{3}k}{(2\pi)^{3}}\frac{1}{2E_{k}^{\pi}}\left[a({\bf k})e^{-ik_{\mathrm{on}}\cdot x}+b^{\dagger}({\bf k})e^{ik_{\mathrm{on}}\cdot x}\right]\;,\label{eq:pi_field}
\end{align}
where $a_{V}$ and $a$ are annihilation operators for $\rho$ and $\pi^{+}$ respectively. Since the $\rho$ meson is charge neutral, $\rho^{\mu}(x)$ is a real vector field. The $\pi^{-}$ field can be obtained by taking the conjugate of $\pi^{+}$. The three spin states of the $\rho$ meson are represented by $\lambda=0,\pm1$ and the polarization four-vector $\epsilon^{\mu}\left(\lambda,p\right)$ is defined as a Lorentz boost of the polarization three-vector $\boldsymbol{\epsilon}_{\lambda}$ in the meson's rest frame~\citep{Sheng:2024kgg}
\begin{align}
\epsilon^{\mu}\left(\lambda,p\right) & =\left(\frac{\mathbf{p}\cdot\boldsymbol{\epsilon}_{\lambda}}{\sqrt{p^{2}}},\boldsymbol{\epsilon}_{\lambda}+\frac{\mathbf{p}\cdot\boldsymbol{\epsilon}_{\lambda}}{\sqrt{p^{2}}\left(p^{0}+\sqrt{p^{2}}\right)}\mathbf{p}\right)\;.\label{eq:epsilon}
\end{align}
We assume that the spin quantization direction is chosen to be along the $y$ direction, so $\boldsymbol{\epsilon}_{\lambda}$ is given by
\begin{align}
\boldsymbol{\epsilon}_{0} & =\left(0,1,0\right)\;,\nonumber \\
\boldsymbol{\epsilon}_{\pm1} & =\mp\frac{1}{\sqrt{2}}\left(\pm i,0,1\right)\;.\label{eq:epsilon_rest}
\end{align}
From the definition given by Eq.~(\ref{eq:epsilon}), one can check the following identities 
\begin{align}
p^{\mu}\epsilon_{\mu}\left(\lambda,p\right) & =0\;,\nonumber \\
\epsilon^{\mu}\left(\lambda_{1},p\right)\epsilon_{\mu}^{*}\left(\lambda_{2},p\right) & =-\delta_{\lambda_{1}\lambda_{2}}\;,\nonumber \\
\underset{\lambda=0,\pm1}{\sum}\epsilon^{\mu*}\left(\lambda,p\right)\epsilon^{\nu}\left(\lambda,p\right) & =-\Delta_{p}^{\mu\nu}\;,
\end{align}
where $\Delta_{p}^{\mu\nu}=g^{\mu\nu}-p^{\mu}p^{\nu}/p^{2}$ denotes the projector perpendicular to $p^{\mu}$. The pion propagator $S^{CTP}\left(x_{1},x_{2}\right)$ is given in~\citep{Yin:2025gvl} and its phase-space form is given by 
\begin{align}
S^{F/\overline{F}}\left(X,p\right) & =\frac{\pm i}{p^{2}-m_{\pi}^{2}\pm i\epsilon}+\frac{\pi}{E_{p}^{\pi}}\delta\left(p^{0}-E_{p}^{\pi}\right)f_{\pi}^{(+)}\left(X,p\right)+\frac{\pi}{E_{p}^{\pi}}\delta\left(p^{0}+E_{p}^{\pi}\right)f_{\pi}^{(-)}\left(X,-p\right)\;,\nonumber \\
S^{</>}\left(X,p\right) & =\frac{\pi}{E_{p}^{\pi}}\delta\left(p^{0}\pm E_{p}^{\pi}\right)+\frac{\pi}{E_{p}^{\pi}}\delta\left(p^{0}-E_{p}^{\pi}\right)f_{\pi}^{(+)}\left(X,p\right)+\frac{\pi}{E_{p}^{\pi}}\delta\left(p^{0}+E_{p}^{\pi}\right)f_{\pi}^{(-)}\left(X,-p\right)\;,\label{eq:S_pi}
\end{align}
where $E_{p}^{\pi}=\sqrt{p^{2}+m_{\pi}^{2}}$ denotes the on-shell energy of pion. The difference between $f^{(+)}$ and $f^{(-)}$ arises from the electric chemical potential and can be neglected. Therefore, the distribution $f_{\pi}^{(+/-)}$ are defined as
\begin{align}
f_{\pi}^{(+/-)}\left(X,p\right) & =\frac{1}{e^{\beta(X)\cdot p}-1}\;,\label{eq:f_pi}
\end{align}
where $\beta^{\mu}\left(X\right)=\beta_{0}\left(X\right)u^{\mu}\left(X\right)$ denotes the thermal velocity. A similar calculation has been performed in Appendix B of~\citep{Dong:2023cng} with a specific choice $u^{\mu}=\left(1,\mathbf{0}\right)$. Since self-energy can be expressed in terms of Lorentz boost invariant quantities, namely $u\cdot p,p^{2}$ and $\beta^{2}$, it is convenient to compute it first in the local rest frame and then boost it into the lab frame. Under the Wigner transformation of Eq.~(\ref{eq:sigma_F}), the leading term of self-energy is 
\begin{align}
\Sigma_{\alpha\beta}^{R(\text{L})}= & g_{\rho\pi}^{2}\int\frac{dp_{1}}{(2\pi)^{4}}\int\frac{dp_{2}}{(2\pi)^{4}}(2\pi)^{4}\delta\left(p-p_{1}+p_{2}\right)\left(ip_{\alpha}^{1}+ip_{\alpha}^{2}\right)\left(ip_{\beta}^{2}+ip_{\beta}^{1}\right)\nonumber \\
 & \times\left[S^{F}\left(X,p_{1}\right)S^{F}\left(X,p_{2}\right)-S^{<}\left(X,p_{1}\right)S^{>}\left(X,p_{2}\right)\right]\nonumber \\
 & +2ig_{\rho\pi}^{2}g_{\alpha\beta}\int\frac{dp_{1}}{(2\pi)^{4}}\int\frac{dp_{2}}{(2\pi)^{4}}(2\pi)^{4}\delta\left(p-p_{1}+p_{2}\right)S^{F}\left(X,p_{1}\right)+\Sigma_{\alpha\beta}^{\text{ren}}\left(p\right)\nonumber \\
= & \Sigma_{\alpha\beta}^{\text{med}}+\Sigma_{\alpha\beta}^{\text{vac}}\;,
\end{align}
where the self-energy is decomposed into medium and vacuum parts. The former contains all $f_{\pi}$-related term while the latter is $f_{\pi}$-independent. The medium contribution evaluates to
\begin{align}
\Sigma_{\alpha\beta}^{\text{med}}= & -g_{\rho\pi}^{2}\int\frac{dp_{1}}{(2\pi)^{4}}\left(2p_{\alpha}^{1}-p_{\alpha}\right)\left(2p_{\beta}^{1}-p_{\beta}\right)\frac{i}{\left(p-p_{1}\right)^{2}-m_{\pi}^{2}+i\left(p^{0}-p_{1}^{0}\right)\epsilon}\nonumber \\
 & \times\frac{\pi}{E_{p_{1}}}\delta\left(p_{1}^{0}-E_{p_{1}}^{\pi}\right)\left[f_{\pi}^{(+)}\left(X,p_{1}\right)+f_{\pi}^{(-)}\left(X,p_{1}\right)\right]\nonumber \\
 & +\left(p\rightarrow-p,\epsilon\rightarrow-\epsilon\right)\nonumber \\
 & +2ig_{\rho\pi}^{2}g_{\alpha\beta}\int \frac{dp_{1}}{(2\pi)^4}\frac{\pi}{E_{p_{1}}^{\pi}}\delta\left(p_{1}^{0}-E_{p_{1}}^{\pi}\right)\left[f_{\pi}^{(+)}\left(X,p_{1}\right)+f_{\pi}^{(-)}\left(X,p_{1}\right)\right]\nonumber \\
= & I_{\alpha\beta}^{1}\left(X,p\right)-I_{\alpha\beta}^{1,*}\left(X,-p\right)+I_{\alpha\beta}^{2}\left(X,p\right)\;.
\end{align}
Because $I_{\alpha\beta}^{1}$ and $I_{\alpha\beta}^{2}$ are Lorentz-boost covariant quantities, they can be computed first in the local rest frame (labeled as $\widetilde{I}_{\alpha\beta}^{1,2}$) and then boosted into lab frame with an identity $I_{\alpha\beta}^{1,2}=\Lambda_{\alpha}^{\;\rho}\Lambda_{\beta}^{\;\sigma}\widetilde{I}_{\rho\sigma}^{1,2}\left(\widetilde{X},\widetilde{p}\right)$. The explicit forms of $\widetilde{I}_{\alpha\beta}^{1}$'s components are computed as 
\begin{align}
\widetilde{I}_{1}^{00}= & \frac{\pi}{|\widetilde{\mathbf{p}}|}\left[4J_{+}\left(2,1\right)-4\widetilde{p}_{0}J_{+}\left(1,1\right)+\widetilde{p}_{0}^{2}J_{+}\left(0,1\right)\right]\;,\nonumber \\
\widetilde{I}_{1}^{0i}= & \frac{\pi p_{i}}{|\widetilde{\mathbf{p}}|^{3}}\left[8|\widetilde{\mathbf{p}}|J_{0}\left(1,2\right)-4|\widetilde{\mathbf{p}}|\widetilde{p}_{0}J_{0}\left(0,2\right)-2\widetilde{p}^{2}J_{+}\left(1,1\right)\right.\nonumber \\
 & \left.+\widetilde{p}_{0}\widetilde{p}^{2}J_{+}\left(0,1\right)+4\widetilde{p}^{0}J_{+}\left(2,1\right)-2\widetilde{p}_{0}^{2}J_{+}\left(1,1\right)\right]\nonumber \\
 & -\frac{\pi\widetilde{p}_{i}}{|\mathbf{\widetilde{p}}|}\left[2J_{+}\left(1,1\right)-\widetilde{p}_{0}J_{+}\left(0,1\right)\right]\;,\nonumber \\
\widetilde{I}_{1}^{ij}= & \pi\frac{\widetilde{p}_{i}\widetilde{p}_{j}}{|\widetilde{\mathbf{p}}|^{5}}\left[-4|\widetilde{\mathbf{p}}|\left(\widetilde{p}_{0}^{2}+|\widetilde{\mathbf{p}}|^{2}\right)J_{0}\left(0,2\right)+8|\widetilde{\mathbf{p}}|\widetilde{p}_{0}J_{0}\left(1,2\right)\right.\nonumber \\
 & \left.+\widetilde{p}_{0}^{4}J_{+}\left(0,1\right)-4\widetilde{p}_{0}^{3}J_{+}\left(1,1\right)+4\widetilde{p}_{0}^{2}J_{+}\left(2,1\right)\right]\nonumber \\
 & +\pi\frac{1}{2|\widetilde{\mathbf{p}}|^{3}}\left(\delta_{ij}-\frac{\widetilde{p}_{i}\widetilde{p}_{j}}{|\widetilde{\mathbf{p}}|^{2}}\right)\left[4|\widetilde{\mathbf{p}}|\widetilde{p}^{2}J_{0}\left(0,2\right)-8|\widetilde{\mathbf{p}}|\widetilde{p}^{0}J_{0}\left(1,2\right)-\widetilde{p}^{4}J_{+}\left(0,1\right)\right.\nonumber \\
 & \left.+4\widetilde{p}^{2}\widetilde{p}_{0}J_{+}\left(1,1\right)-4\widetilde{p}_{0}^{2}J_{+}\left(2,1\right)+4|\widetilde{\mathbf{p}}|^{2}J_{+}\left(0,3\right)\right]\;.
\end{align}
 where the special functions $J_{0,\pm}$ are defined as
\begin{align}
J_{\pm}\left(n_{1},n_{2}\right) & =-ig_{\rho\pi}^{2}\int\frac{d|\mathbf{k}|}{(2\pi)^{3}2E_{k}^{\pi}}E_{\pi,k}^{n_{1}}|\mathbf{k}|^{n_{2}}\frac{2}{e^{\beta E_{k}^{\pi}}-1}\ln\frac{\widetilde{p}^{2}\mp2\widetilde{p}^{0}E_{k}^{\pi}+2|\widetilde{\mathbf{p}}||\mathbf{k}|+i\left(\widetilde{p}^{0}\mp E_{k}^{\pi}\right)\epsilon}{\widetilde{p}^{2}\mp2\widetilde{p}^{0}E_{k}^{\pi}-2|\widetilde{\mathbf{p}}||\mathbf{k}|+i\left(\widetilde{p}^{0}\mp E_{k}^{\pi}\right)\epsilon}\;,\nonumber \\
J_{0}\left(n_{1},n_{2}\right) & =-ig_{\rho\pi}^{2}\int\frac{d|\mathbf{k}|}{(2\pi)^{3}2E_{k}^{\pi}}E_{\pi,k}^{n_{1}}|\mathbf{k}|^{n_{2}}\frac{2}{e^{\beta E_{k}^{\pi}}-1}\;.
\end{align}
For simplicity, the dependence of these special functions on $p^{\mu}$ and $X^{\mu}$ is suppressed. In contrast to $I_{1}^{\alpha\beta}$, the term $I_{2}^{\alpha\beta}$ is proportional to $g^{\alpha\beta}$, so it can be computed in lab frame directly
\begin{align}
I_{\alpha\beta}^{2} & =2ig^{2}g_{\alpha\beta}\int\frac{d^{3}\mathbf{k}}{(2\pi)^{3}2E_{k}^{\pi}}\frac{2}{e^{\beta_{0}E_{k}^{\pi}}-1}=-8\pi g_{\alpha\beta}J_{0}\left(0,2\right)\;.
\end{align}
Before performing the Lorentz boost to $\widetilde{I}_{1}^{\alpha\beta}$, we first introduce some useful identities on the boost matrix $\Lambda_{\;\nu}^{\mu}$. This matrix, mapping the local rest frame with the lab frame, is defined by $\Lambda_{\;\nu}^{\mu}\left(1,\mathbf{0}\right)^{\nu}=u^{\mu}$, which implies
\begin{align}
\Lambda_{\;0}^{\mu} & =u^{\mu},\Lambda_{\;i}^{\mu}\widetilde{p}^{i}=p^{\mu}-\left(u\cdot p\right)u^{\mu}\;,\nonumber \\
\widetilde{p}_{0} & =u\cdot p,|\widetilde{\mathbf{p}}|^{2}=\left(u\cdot p\right)^{2}-p^{2}\;.\label{eq:Lambda}
\end{align}
Using these identities, the medium contribution in lab frame can be rewritten in a standard decomposition
\begin{align}
\Sigma_{\text{med}}^{\mu\nu}= & -\Delta_{L}^{\mu\nu}\Sigma_{\text{med}}^{L}-\Delta_{T}^{\mu\nu}\Sigma_{\text{med}}^{T}\;,
\end{align}
where the medium-induced transverse/longitudinal self-energy is given by
\begin{align}
\Sigma_{\text{med}}^{L}= & \pi\frac{1}{|\widetilde{\mathbf{p}}|^{3}}\left\{ -8|\widetilde{\mathbf{p}}|\widetilde{p}^{2}J_{0}\left(0,2\right)+\widetilde{p}_{0}^{2}\widetilde{p}^{2}\left[J_{+}\left(0,1\right)+J_{-}\left(0,1\right)\right]\right.\nonumber \\
 & \left.-4\widetilde{p}_{0}\widetilde{p}^{2}\left[J_{+}\left(1,1\right)-J_{-}\left(1,1\right)\right]+4\widetilde{p}^{2}\left[J_{+}\left(2,1\right)+J_{-}\left(2,1\right)\right]\right\}\;,\nonumber \\
\Sigma_{\text{med}}^{T}= & \pi\frac{1}{2|\widetilde{\mathbf{p}}|^{3}}\left\{ 8|\widetilde{\mathbf{p}}|\left(\widetilde{p}_{0}^{2}+|\widetilde{\mathbf{p}}|^{2}\right)J_{0}\left(0,2\right)-\widetilde{p}^{4}\left[J_{+}\left(0,1\right)+J_{-}\left(0,1\right)\right]\right.\nonumber \\
 & +4\widetilde{p}^{2}\widetilde{p}_{0}\left[J_{+}\left(1,1\right)-J_{-}\left(1,1\right)\right]-4\widetilde{p}_{0}^{2}\left[J_{+}\left(2,1\right)+J_{-}\left(2,1\right)\right]\nonumber \\
 & \left.+4|\widetilde{\mathbf{p}}|^{2}\left[J_{+}\left(0,3\right)+J_{-}\left(0,3\right)\right]\right\}\;,\label{eq:medium-retarded}
\end{align}
For the vacuum contribution, since it is independent of flow velocity, it can be computed directly in the lab frame. 
After renormalization with the condition $\text{Im}\left[\Sigma^{\text{vac}}\left(p^{2}=m^{2}\right)\right]=0$, the final result reads.
\begin{align}
\Sigma_{\alpha\beta}^{\text{vac}}= & \frac{\pi g^{2}}{(4\pi)^{2}}p^{2}\Delta_{\alpha\beta}^{p}\int_{0}^{1} dx\left(2x-1\right)^{2}\theta\left(-x^{2}p^{2}+xp^{2}-m_{\pi}^{2}\right)\nonumber \\
 & +i\frac{g^{2}}{(4\pi)^{2}}p^{2}\Delta_{\alpha\beta}^{p}\int_{0}^{1} dx\left(2x-1\right)^{2}\log\frac{|x^{2}p^{2}-xp^{2}+m_{\pi}^{2}|}{|x^{2}m_{V}^{2}-xm_{V}^{2}+m_{\pi}^{2}|}\nonumber \\
= & \Delta_{\alpha\beta}^{p}\Sigma_{\text{vac}}\;.\label{eq:vaccum-retarded}
\end{align}
Combining the medium and vacuum terms together, we obtain the full retarded self-energy 
\begin{align}
\Sigma_{R}^{\mu\nu(\text{L})}\left(X,p\right) & =-\Delta_{L}^{\mu\nu}\Sigma_{L}-\Delta_{T}^{\mu\nu}\Sigma_{T}\;,
\end{align}
where transverse/longitudinal self-energy is given by
\begin{align}
\Sigma^{L/T}= & \Sigma_{\text{med}}^{L/T}-\Sigma_{\text{vac}}\;.\label{eq:sigma_L/T}
\end{align}
The previous calculation shows explicitly that the transverse and longitudinal parts of self-energy are functions of $p^{2},\sqrt{\beta^{2}},u\cdot p$, ensuring that $\Sigma^{L/T}$ is Lorentz boost invariant. 

The less self-energy can be obtained following the same method and it is expressed in compact form 
\begin{align}
\Sigma_{<,\text{L}}^{\alpha\beta} & =-\Delta_{L}^{\alpha\beta}\Sigma_{<}^{L}-\Delta_{T}^{\alpha\beta}\Sigma_{<}^{T}\;,
\end{align}
where 
\begin{align}
\Sigma_{L}^{<} & =\frac{p^2}{2|\widetilde{\mathbf{p}}|^{3}}\left[4M_{0}\left(2,1\right)-4\widetilde{p}_{0}M_{0}\left(1,1\right)+\widetilde{p}_{0}^{2}M_{0}\left(0,1\right)\right]\;,\nonumber \\
\Sigma_{T}^{<} & =\frac{1}{4|\widetilde{\mathbf{p}}|^{3}}\left[4|\widetilde{\mathbf{p}}|^{2}M_{0}\left(0,3\right)-4\widetilde{p}_{0}^{2}M_{0}\left(2,1\right)+4\widetilde{p}_{0}\widetilde{p}^{2}M_{0}\left(1,1\right)-\widetilde{p}^{4}M_{0}\left(0,1\right)\right]\;.
\end{align}
The special function of $M_{0}\left(n_{1},n_{2}\right)$ is defined as
\begin{align}
M_{0}\left(n_{1},n_{2}\right) & =-g_{\rho\pi}^{2}\int_{k_{min}}^{k_{max}}\frac{d|\mathbf{k}|}{(2\pi)2E_{k}^{\pi}}E_{\pi,k}^{n_{1}}|\mathbf{k}|^{n_{2}}\frac{1}{e^{\beta_{0}E_{k}^{\pi}}-1}\frac{1}{e^{\beta_{0}\left(\widetilde{p}_{0}-E_{k}^{\pi}\right)}-1}\;,
\end{align}
where the exact value of the integration limits $k_{min}$ and $k_{max}$ reads
\begin{align}
k_{min}=\left|-\frac{|\widetilde{\mathbf{p}}|}{2}+\widetilde{p}_{0}\sqrt{\frac{1}{4}-\frac{m_\pi^{2}}{p^{2}}}\right|,\; & k_{max}=\frac{|\widetilde{\mathbf{p}}|}{2}+\widetilde{p}_{0}\sqrt{\frac{1}{4}-\frac{m_\pi^{2}}{p^{2}}}\;.
\end{align}

\section{Calculation of next-to-leading-order self-energies\label{appendix:self-energy-correction}}

In this section, we present the detailed calculation of the NLO self-energies. 
As discussed in Appendix~\ref{appendix:calculation-of-self}, to calculate the self-energy in the lab frame, we can start with the local rest frame and then perform a Lorentz boost. The calculation in this section follows the same procedure.

We begin with the definition of NLO self-energy in phase space which is given by a Wigner transformation. 
\begin{align}
\Sigma_{\alpha\beta}^{R(\text{NL})}= & -g_{\rho\pi}^{2}\int\frac{dp_{1}}{(2\pi)^{4}}\int\frac{dp_{2}}{(2\pi)^{4}}(2\pi)^{4}\delta\left(p-p_{1}+p_{2}\right)\nonumber \\
 & \times\left[\partial_{[\alpha}^{X}S^{F}\left(X,p_{1}\right)\left(ip_{\beta]}^{2}+ip_{\beta]}^{1}\right)S^{F}\left(X,p_{2}\right)-\partial_{[\alpha}^{X}S^{F}\left(X,p_{2}\right)\left(ip_{\beta]}^{1}+ip_{\beta]}^{2}\right)S^{F}\left(X,p_{1}\right)\right.\nonumber \\
 & \left.-\partial_{[\alpha}^{X}S^{<}\left(X,p_{1}\right)\left(ip_{\beta]}^{2}+ip_{\beta]}^{1}\right)S^{>}\left(X,p_{2}\right)+\partial_{[\alpha}^{X}S^{>}\left(X,p_{2}\right)\left(ip_{\beta]}^{1}+ip_{\beta]}^{2}\right)S^{<}\left(X,p_{1}\right)\right]\;,\label{eq:sigma_R_NL}\\
\Sigma_{\alpha\beta}^{<(\text{NL})}= & -g_{\rho\pi}^{2}\int\frac{dp_{1}}{(2\pi)^{4}}\int\frac{dp_{2}}{(2\pi)^{4}}(2\pi)^{4}\delta\left(p-p_{1}+p_{2}\right)\nonumber \\
 & \times\left[\partial_{[\alpha}^{X}S^{<}\left(X,p_{1}\right)\left(ip_{\beta]}^{2}+ip_{\beta]}^{1}\right)S^{>}\left(X,p_{2}\right)-\partial_{[\alpha}^{X}S^{>}\left(X,p_{2}\right)\left(ip_{\beta]}^{1}+ip_{\beta]}^{2}\right)S^{<}\left(X,p_{1}\right)\right]\;.\label{eq:sigma_<_NL}
\end{align}
Inserting the pion distribution function $f_{\pi}$ (Eq.~(\ref{eq:f_pi})) into Eq.~(\ref{eq:sigma_<_NL}), we obtain the following
\begin{align}
\Sigma_{\alpha\beta}^{<(\text{NL})}= & -4ig_{\rho\pi}^{2}\partial_{[\alpha}^{X}\beta^{\sigma}\int\frac{dp_{1}}{(2\pi)^{4}}\int\frac{dp_{2}}{(2\pi)^{4}}(2\pi)^{4}\delta\left(p-p_{1}-p_{2}\right)p_{\beta]}^{1}p_{\sigma}^{1}\nonumber \\
 & \times\frac{\pi}{E_{p_{1}}^{\pi}}\delta\left(p_{1}^{0}-E_{p_{1}}^{\pi}\right)\partial_{\beta\cdot p_{1}}f_{\pi}^{(+)}\left(X,p_{1}\right)\frac{\pi}{E_{p_{2}}^{\pi}}\delta\left(p_{2}^{0}-E_{p_{2}}^{\pi}\right)f_{\pi}^{(-)}\left(X,p_{2}\right)\nonumber \\
 & +2p_{[\alpha}H^{\prime}_{4,\beta]}\nonumber \\
= & -2\partial_{[\alpha}^{X}\beta^{\sigma}\Lambda_{\beta]}^{\;\rho}\Lambda_{\sigma}^{\;\gamma}\widetilde{Q}_{\rho\gamma}+2p_{[\alpha}H^{\prime}_{4,\beta]}\;,\label{eq:self_energy_cal}
\end{align}
where $\Lambda_{\;\nu}^{\mu}$ is the boost matrix defined in the Appendix ~\ref{appendix:calculation-of-self}. $H^\prime_{4,\beta}$ is a function of $x^\mu$ and $p^\mu$. The tensor $\widetilde{Q}^{\mu\nu}$ is defined in the local rest frame and its explicit form is given by
\begin{align}
\widetilde{Q}^{\mu\nu}= & 2ig^{2}\int\frac{d|\mathbf{k}|d\cos\phi d\theta}{(2\pi)^{3}2E_{k}^{\pi}2|\widetilde{\mathbf{p}}|}|\mathbf{k}|2\pi\delta\left(\cos\phi-\frac{2\widetilde{p}_{0}E_{k}^{\pi}-\widetilde{p}^{2}}{2|\widetilde{\mathbf{p}}||\mathbf{k}|}\right)k^{\mu}k^{\nu}\nonumber \\
 & \times\left(\partial_{\beta_{0}E_{k}^{\pi}}\frac{1}{e^{\beta_{0}E_{k}^{\pi}}-1}\right)\frac{1}{e^{\beta_{0}\widetilde{p}^{0}-\beta_{0}E_{k}^{\pi}}-1}\;.
\end{align}
The values of each component of $\widetilde{Q}^{\rho\sigma}$ are given by
\begin{align}
\widetilde{Q}^{00}= & \frac{ig_{\rho\pi}^{2}}{|\widetilde{\mathbf{p}}|}\int\frac{d|\mathbf{k}|}{(2\pi)2E_{k}^{\pi}}|\mathbf{k}|E_{\pi,k}^{2}\left(\partial_{\beta_{0}E_{k}^{\pi}}\frac{1}{e^{\beta_{0}E_{k}^{\pi}}-1}\right)\frac{1}{e^{\beta_{0}\widetilde{p}^{0}-\beta_{0}E_{k}^{\pi}}-1}\nonumber \\
= & \frac{1}{|\widetilde{\mathbf{p}}|}L_{0}\left(2,1\right)=A\;,\nonumber \\
\widetilde{Q}^{0i}= & ig^{2}\frac{\widetilde{\mathbf{p}}^{i}}{2|\widetilde{\mathbf{p}}|^{3}}\int\frac{d|\mathbf{k}|}{(2\pi)2E_{k}^{\pi}}|\mathbf{k}|E_{k}^{\pi}\left(2\widetilde{p}_{0}E_{k}^{\pi}-\widetilde{p}^{2}\right)\left(\partial_{\beta_{0}E_{k}^{\pi}}\frac{1}{e^{\beta_{0}E_{k}^{\pi}}-1}\right)\frac{1}{e^{\beta_{0}\widetilde{p}^{0}-\beta_{0}E_{k}^{\pi}}-1}\nonumber \\
= & \frac{\widetilde{\mathbf{p}}^{i}}{2|\widetilde{\mathbf{p}}|^{3}}\left[2\widetilde{p}_{0}L_{0}\left(2,1\right)-\widetilde{p}^{2}L_{0}\left(1,1\right)\right]=\frac{\widetilde{\mathbf{p}}^{i}}{|\widetilde{\mathbf{p}}|}B\;,\nonumber \\
\widetilde{Q}^{ij}= & \frac{\widetilde{\mathbf{p}}^{i}\widetilde{\mathbf{p}}^{j}}{4|\mathbf{p}|^{5}}\left[4\widetilde{p}_{0}^{2}L_{0}\left(2,1\right)-4\widetilde{p}_{0}p^{2}L_{0}\left(1,1\right)+\widetilde{p}^{4}L_{0}\left(0,1\right)\right]\nonumber \\
 & +\left(\delta^{ij}-\frac{\widetilde{\mathbf{p}}^{i}\widetilde{\mathbf{p}}^{j}}{|\widetilde{\mathbf{p}}|^{2}}\right)\frac{1}{8|\widetilde{\mathbf{p}}|^{3}}\left[-4\widetilde{p}_{0}^{2}L_{0}\left(2,1\right)+4\widetilde{p}_{0}p^{2}L_{0}\left(1,1\right)-\widetilde{p}^{4}L_{0}\left(0,1\right)+4|\widetilde{\mathbf{p}}|^{2}L_{0}\left(0,3\right)\right]\nonumber \\
= & \frac{\widetilde{\mathbf{p}}^{i}\widetilde{\mathbf{p}}^{j}}{|\widetilde{\mathbf{p}}|^{2}}C+\left(\delta^{ij}-\frac{\widetilde{\mathbf{p}}^{i}\widetilde{\mathbf{p}}^{j}}{|\widetilde{\mathbf{p}}|^{2}}\right)D\;.
\end{align}
By introducing a Lorentz boost, the tensor $Q^{\mu\nu}$ in the lab frame is given by
\begin{align}
Q^{\rho\sigma}= & \frac{u^{\rho}p^{\sigma}\left(u\cdot p\right)}{\left(u\cdot p\right)^{2}-p^{2}}\left(\frac{\sqrt{\left(u\cdot p\right)^{2}-p^{2}}}{u\cdot p}B-C+D\right)-g^{\rho\sigma}D\nonumber \\
 & +u^{\rho}u^{\sigma}\left(A-\frac{2\left(u\cdot p\right)}{\sqrt{\left(u\cdot p\right)^{2}-p^{2}}}B+\frac{\left(u\cdot p\right)^{2}}{\left(u\cdot p\right)^{2}-p^{2}}C-\frac{p^{2}}{\left(u\cdot p\right)^{2}-p^{2}}D\right)\nonumber \\
 & +p^{\rho}H^{\prime\prime,\sigma}_{4}\;.\label{eq:I}
\end{align}
Inserting Eq.~(\ref{eq:I}) into Eq.~(\ref{eq:self_energy_cal}), the self-energy correction can be rewritten as
\begin{align}
\Sigma_{\alpha\beta}^{<\left(\text{NL}\right)}= & -2\partial_{[\alpha}^{X}\beta_{\beta]}H_{1}-2u_{[\beta}\partial_{\alpha]}^{X}\beta_{\rho}p^{\rho}H_{2}\nonumber \\
 & -2u_{[\beta}\partial_{\alpha]}^{X}\beta_{\rho}u^{\rho}H_{3}+2p_{[\alpha}H_{4,\beta]}\;,
\end{align}
where the coefficient functions $H_{i=1,2,3}$ are given by
\begin{align}
H_{1}= & \frac{1}{8|\widetilde{\mathbf{p}}|^{3}}\left[4\widetilde{p}_{0}^{2}L_{0}\left(2,1\right)-4\widetilde{p}_{0}p^{2}L_{0}\left(1,1\right)+\widetilde{p}^{4}L_{0}\left(0,1\right)-4|\widetilde{\mathbf{p}}|^{2}L_{0}\left(0,3\right)\right]\;,\nonumber \\
H_{2}= & \frac{1}{8|\widetilde{\mathbf{p}}|^{5}}\left\{ \left[-4\widetilde{p}_{0}\left(2\widetilde{p}^{2}+\widetilde{p}_{0}^{2}\right)L_{0}\left(2,1\right)+4\widetilde{p}^{2}\left(\widetilde{p}^{2}+2\widetilde{p}_{0}^{2}\right)L_{0}\left(1,1\right)\right]\right.\nonumber \\
 & \left.-\widetilde{p}_{0}\left[3\widetilde{p}^{4}L_{0}\left(0,1\right)-4|\widetilde{\mathbf{p}}|^{2}L_{0}\left(0,3\right)\right]\right\}\;,\nonumber \\
H_{3}= & \frac{1}{8|\widetilde{\mathbf{p}}|^{5}}\left\{ 4p^{2}\left(2\widetilde{p}^{2}+\widetilde{p}_{0}^{2}\right)L_{0}\left(2,1\right)-12\widetilde{p}_{0}p^{4}L_{0}\left(1,1\right)\right.\nonumber \\
 & \left.+\widetilde{p}^{4}\left(2\widetilde{p}_{0}^{2}+p^{2}\right)L_{0}\left(0,1\right)-4p^{2}|\widetilde{\mathbf{p}}|^{2}L_{0}\left(0,3\right)\right\}\;,\label{eq:E}
\end{align}
with the special function $L_{0}$ defined as
\begin{align}
L_{0}\left(n_{1},n_{2}\right)= & ig_{\rho\pi}^{2}\int_{k_{min}}^{k_{max}}\frac{d|\mathbf{k}|}{(2\pi)2E_{k}^{\pi}}E_{\pi,k}^{n_{1}}|\mathbf{k}|^{n_{2}}\left(\partial_{\beta_{0}E_{k}^{\pi}}\frac{1}{e^{\beta_{0}E_{k}^{\pi}}-1}\right)\frac{1}{e^{\beta_{0}\widetilde{p}^{0}-\beta_{0}E_{k}^{\pi}}-1}\;,\label{eq:L}
\end{align}
Applying the same procedure to the retarded self-energy correction gives the explicit form of coefficient functions $F_{i}$ 
\begin{align}
F_{1}= & \frac{\pi}{|\widetilde{\mathbf{p}}|^{3}}\left\{ -4|\widetilde{\mathbf{p}}|\widetilde{p}^{0}K_{0}\left(1,2\right)-\frac{1}{4}\widetilde{p}^{4}\left[K_{+}\left(0,1\right)-K_{-}\left(0,1\right)\right]\right.\nonumber \\
 & +\widetilde{p}^{2}\widetilde{p}^{0}\left[K_{+}\left(1,1\right)+K_{-}\left(1,1\right)\right]-\widetilde{p}_{0}^{2}\left[K_{+}\left(2,1\right)-K_{-}\left(2,1\right)\right]\nonumber \\
 & \left.+|\widetilde{\mathbf{p}}|^{2}\left[K_{+}\left(0,3\right)-K_{-}\left(0,3\right)\right]\right\}\;,\nonumber \\
F_{2}= & \frac{\pi}{|\widetilde{\mathbf{p}}|^{5}}\left\{ |\widetilde{\mathbf{p}}|\left(4\widetilde{p}_{0}^{2}+8\widetilde{p}^{2}\right)K_{0}\left(1,2\right)+\frac{3}{4}\widetilde{p}^{0}\widetilde{p}^{4}\left[K_{+}\left(0,1\right)-K_{-}\left(0,1\right)\right]\right.\nonumber \\
 & -\widetilde{p}^{2}\left(\widetilde{p}^{2}+2\widetilde{p}_{0}^{2}\right)\left[K_{+}\left(1,1\right)+K_{-}\left(1,1\right)\right]+\left(\widetilde{p}_{0}^{2}+2\widetilde{p}^{2}\right)\widetilde{p}^{0}\left[K_{+}\left(2,1\right)-K_{-}\left(2,1\right)\right]\nonumber \\
 & \left.-|\widetilde{\mathbf{p}}|^{2}\widetilde{p}^{0}\left[K_{+}\left(0,3\right)-K_{-}\left(0,3\right)\right]\right\}\;,\nonumber \\
F_{3}= & \frac{\pi}{|\widetilde{\mathbf{p}}|^{5}}\left\{ 4|\widetilde{\mathbf{p}}|\widetilde{p}^{0}\left(2|\widetilde{\mathbf{p}}|^{2}-3\widetilde{p}^{2}\right)K_{0}\left(1,2\right)-\left(2\widetilde{p}^{4}+\widetilde{p}^{2}\widetilde{p}_{0}^{2}\right)\left[K_{+}\left(2,1\right)-K_{-}\left(2,1\right)\right]\right.\nonumber \\
 & +3\widetilde{p}^{4}\widetilde{p}^{0}\left[K_{+}\left(1,1\right)+K_{-}\left(1,1\right)\right]-\frac{1}{4}\widetilde{p}^{4}\left(2\widetilde{p}_{0}^{2}+\widetilde{p}^{2}\right)\left[K_{+}\left(0,1\right)-K_{-}\left(0,1\right)\right]\nonumber \\
 & \left.+\widetilde{p}^{2}|\widetilde{\mathbf{p}}|^{2}\left[K_{+}\left(0,3\right)-K_{-}\left(0,3\right)\right]\right\}\;,\label{eq:F1-3}
\end{align}
where the special functions $K_{0,\pm}$ are given by
\begin{align}
K_{0}\left(n_{1},n_{2}\right)= & \frac{1}{2}g_{\rho\pi}^{2}\int\frac{d|\mathbf{k}|}{(2\pi)^{3}2E_{k}^{\pi}}E_{\pi,k}^{n_{1}}|\mathbf{k}|^{n_{2}}\partial_{\beta_{0}E_{k}^{\pi}}\frac{2}{e^{\beta_{0}E_{k}^{\pi}}-1}\;,\nonumber \\
K_{\pm}\left(n_{1},n_{2}\right)= & \frac{1}{2}g_{\rho\pi}^{2}\int\frac{d|\mathbf{k}|}{(2\pi)^{3}2E_{k}^{\pi}}E_{\pi,k}^{n_{1}}|\mathbf{k}|^{n_{2}}\partial_{\beta_{0}E_{k}^{\pi}}\frac{2}{e^{\beta_{0}E_{k}^{\pi}}-1}\nonumber \\
 & \times\ln\frac{\widetilde{p}^{2}\mp2\widetilde{p}^{0}E_{k}^{\pi}+2|\widetilde{\mathbf{p}}||\mathbf{k}|+i\left(\widetilde{p}^{0}\mp E_{k}^{\pi}\right)\epsilon}{\widetilde{p}^{2}\mp2\widetilde{p}^{0}E_{k}^{\pi}-2|\widetilde{\mathbf{p}}||\mathbf{k}|+i\left(\widetilde{p}^{0}\mp E_{k}^{\pi}\right)\epsilon}\;,\label{eq:K}
\end{align}

\section{Spin density matrix\label{appendix:spin-density-matrix}}

In this section, we calculate the diagonal elements of the spin density matrix.
To simplify the calculation, we first introduce several useful identities for the projector operators
\begin{align}
\Delta_{T}^{\mu\nu}u_{\nu}^{L} & =0\;,\nonumber \\
\Delta_{L}^{\mu\nu}u_{\nu}^{L} & =u_{L}^{\mu}\;,\nonumber \\
\Delta_{a}^{\mu\rho}\Delta_{\rho\nu}^{b} & =\Delta_{a,\nu}^{\mu}\delta_{ab}\;,\nonumber \\
\partial_{X}^{\gamma}\Delta_{L}^{\mu\nu} & =-\partial_{X}^{\gamma}\Delta_{T}^{\mu\nu}=\frac{1}{u_{L}^{2}}\partial_{X}^{\gamma}u^{\tau}\left(\Delta_{\tau}^{T,\mu}u_{L}^{\nu}+\Delta_{\tau}^{T,\nu}u_{L}^{\mu}\right)\;,\nonumber \\
\partial_{p}^{\gamma}\Delta_{L}^{\mu\nu} & =-\frac{u\cdot p}{p^{2}u_{L}^{2}}\left(\Delta_{T}^{\mu\gamma}u_{L}^{\nu}+\Delta_{T}^{\nu\gamma}u_{L}^{\mu}\right)-\frac{1}{p^{2}}\left(p^{\mu}\Delta_{L}^{\gamma\nu}+p^{\nu}\Delta_{L}^{\gamma\mu}\right)\;,\nonumber \\
\partial_{p}^{\gamma}\Delta_{T}^{\mu\nu} & =\frac{u\cdot p}{p^{2}u_{L}^{2}}\left(\Delta_{T}^{\mu\gamma}u_{L}^{\nu}+\Delta_{T}^{\nu\gamma}u_{L}^{\mu}\right)-\frac{1}{p^{2}}\left(p^{\mu}\Delta_{T}^{\gamma\nu}+p^{\nu}\Delta_{T}^{\gamma\mu}\right)\;,\label{eq:gradient_Delta}
\end{align}
where the longitudinal flow velocity is defined as $u_L^\mu=\Delta^{\mu\nu}_p u_\nu$ and its spatial components are parallel to $\mathbf{p}$ in the local rest frame. 
The contribution from self-energy correction can be obtained by inserting the Eqs. (\ref{eq:sigma_R_NL},\ref{eq:sigma_<_NL}) into Eq.~(\ref{eq:G_NL_<_solution}).
The corresponding expressions can be obtained by direct substitution and are collected in Table.~\ref{tab:h_i}. 
The contribution from the Poisson bracket is given by
\begin{align}
f_{\lambda_{1}\lambda_{1}}^{\text{poi}}= & -2\text{Im}\left[\epsilon_{\lambda_{1}}^{\mu,*}\epsilon_{\lambda_{1}}^{\nu}\Delta_{\mu\rho}^{a}\rho_{a}\partial_{X}^{\gamma}\left(\Delta_{b}^{\rho\sigma}\frac{1}{\rho_{b}}\right)\partial_{\gamma}^{p}\left(n_{B}\Delta_{\sigma\nu}^{c}\text{Re}\left(i\rho_{c}^{*}\right)\right)\right]\nonumber \\
 & +2\text{Im}\left[\epsilon_{\lambda_{1}}^{\mu,*}\epsilon_{\lambda_{1}}^{\nu}\Delta_{\mu\rho}^{a}\rho_{a}\partial_{\gamma}^{p}\left(\Delta_{b}^{\rho\sigma}\frac{1}{\rho_{b}}\right)\partial_{X}^{\gamma}\left(n_{B}\Delta_{\sigma\nu}^{c}\text{Re}\left(i\rho_{c}^{*}\right)\right)\right]\nonumber \\
 & -2\text{Im}\left[\epsilon_{\lambda_{1}}^{\mu,*}\epsilon_{\lambda_{1}}^{\nu}\partial_{X}^{\gamma}\left(\Delta_{\mu\rho}^{a}\rho_{a}\right)\Delta_{b}^{\rho\sigma}n_{B}\text{Re}\left(\frac{i}{\rho_{b}}\right)\partial_{\gamma}^{p}\left(\Delta_{\sigma\nu}^{c}\rho_{c}^{*}\right)\right]\;.
\end{align}
For brevity, we do not show the full derivation for all terms. Instead, we illustrate the procedure with one example
\begin{align}
f_{\lambda_{1}\lambda_{1}}^{\text{ex}}= & -2\text{Im}\left[\epsilon_{\lambda_{1}}^{\mu,*}\epsilon_{\lambda_{1}}^{\nu}\partial_{X}^{\gamma}\left(\Delta_{\mu\rho}^{a}\right)\rho_{a}\Delta_{b,\nu}^{\rho}n_{B}\text{Re}\left(\frac{i}{\rho_{b}}\right)\partial_{\gamma}^{p}\left(\rho_{b}^{*}\right)\right]\nonumber \\
= & -2\text{Im}\left[\epsilon_{\lambda_{1}}^{\mu,*}\epsilon_{\lambda_{1}}^{\nu}\frac{1}{u_{L}^{2}}\partial_{X}^{\gamma}u^{\tau}\Delta_{\tau\mu}^{T}u_{\rho}^{L}\left(\rho_{L}-\rho_{T}\right)\Delta_{b,\nu}^{\rho}n_{B}\text{Re}\left(\frac{i}{\rho_{b}}\right)\partial_{\gamma}^{p}\left(\rho_{b}^{*}\right)\right]\nonumber \\
 & -2\text{Im}\left[\epsilon_{\lambda_{1}}^{\mu,*}\epsilon_{\lambda_{1}}^{\nu}\frac{1}{u_{L}^{2}}\partial_{X}^{\gamma}u^{\tau}\Delta_{\tau\rho}^{T}u_{\mu}^{L}\left(\rho_{L}-\rho_{T}\right)\Delta_{b,\nu}^{\rho}n_{B}\text{Re}\left(\frac{i}{\rho_{b}}\right)\partial_{\gamma}^{p}\left(\rho_{b}^{*}\right)\right]\nonumber \\
= & -2\partial_{X}^{\gamma}u^{\tau}\text{Im}\left[\epsilon_{\lambda_{1}}^{\mu,*}\epsilon_{\lambda_{1}}^{\nu}\Delta_{\mu\tau}^{T}u_{\gamma}u_{\nu}\frac{1}{u_{L}^{2}}\left(\rho_{L}-\rho_{T}\right)n_{B}\text{Re}\left(\frac{i}{\rho_{L}}\right)\partial_{u\cdot p}\left(\rho_{L}^{*}\right)\right]\nonumber \\
 & -4\partial_{X}^{\gamma}u^{\tau}\text{Im}\left[\epsilon_{\lambda_{1}}^{\mu,*}\epsilon_{\lambda_{1}}^{\nu}\Delta_{\mu\tau}^{T}p_{\gamma}u_{\nu}\frac{1}{u_{L}^{2}}\left(\rho_{L}-\rho_{T}\right)n_{B}\text{Re}\left(\frac{i}{\rho_{L}}\right)\partial_{p^{2}}\left(\rho_{L}^{*}\right)\right]\nonumber \\
 & -2\partial_{X}^{\gamma}u^{\tau}\text{Im}\left[\epsilon_{\lambda_{1}}^{\nu,*}\epsilon_{\lambda_{1}}^{\mu}\Delta_{\mu\tau}^{T}u_{\gamma}u_{\nu}\frac{1}{u_{L}^{2}}\left(\rho_{L}-\rho_{T}\right)n_{B}\text{Re}\left(\frac{i}{\rho_{T}}\right)\partial_{u\cdot p}\left(\rho_{T}^{*}\right)\right]\nonumber \\
 & -4\partial_{X}^{\gamma}u^{\tau}\text{Im}\left[\epsilon_{\lambda_{1}}^{\nu,*}\epsilon_{\lambda_{1}}^{\mu}\Delta_{\mu\tau}^{T}p_{\gamma}u_{\nu}\frac{1}{u_{L}^{2}}\left(\rho_{L}-\rho_{T}\right)n_{B}\text{Re}\left(\frac{i}{\rho_{T}}\right)\partial_{p^{2}}\left(\rho_{T}^{*}\right)\right]\;.
\end{align}
This term will contribute to $h_{7,8,9,10}$ and remaining terms can be calculated in the same way. 
The resulting terms can be classified by different tensor structures and their corresponding coefficients $h_i$ are collected in the Table.~\ref{tab:h_i}. The first 6 components, i.e. $h_{1,\cdots,6}$, denote the contribution from self-energy correction while the rest 11 components, i.e. $h_{7,\cdots,17}$, denote contribution from Poisson bracket. 
With the explicit form of self-energy, one can verify that the imaginary parts of $h_i$ are not independent and some useful identities are given by
\begin{align}
    2\text{Im}\left(h_{1}-h_{2}\right)-\frac{(u\cdot p)^2-p^2}{p^2}\text{Im}\left(h_{5}-h_{6}\right)&=\frac{2}{\beta_{0}}\text{Im}\left(h_{11}+h_{12}\right)\;,\nonumber\\
    p^2\text{Im}\left(h_{3}-h_{4}\right)+(u\cdot p) \text{Im}\left(h_{5}-h_{6}\right)&=0\;,\nonumber\\
\text{Im}\left(h_{7}+h_{8}\right)&=\text{Im}\left(h_{11}+h_{12}\right)\;,\nonumber\\
\text{Im}\left(h_{9}+h_{10}\right)&=0\;,\nonumber\\
\text{Im}\left(h_{13}+h_{14}\right)&=\frac{u\cdot p}{2\beta_{0}^{2}}\text{Im}\left(h_{11}+h_{12}\right)\;,\nonumber\\
\frac{u\cdot p}{2\beta_{0}^{2}}\text{Im}h_{15}^{a}&=\text{Im}h_{17}^{a}\;,
\end{align}
Then, the special functions $\mathcal{T}$ and $\mathcal{N}$ can be simplified into  
\begin{align}
\mathcal{T}^{0}= & 2\int_0^{\infty}\frac{dp^{0}}{\pi}p^{0}n_{B}\left[\text{Im}\left(\rho_{T}^{*}\right)-\epsilon_{0}^{\mu}\epsilon_{0}^{\nu}\Delta_{\mu\nu}^{L}\text{Im}\left(\rho_{L}^{*}-\rho_{T}^{*}\right)\right]\;,\nonumber\\
\mathcal{T}_{\rho}^{1}= &\frac{2}{\beta_{0}}\int_0^{\infty}\frac{dp^{0}}{\pi}p^{0}\Delta_{\mu\rho}^{T}u_{\nu}\epsilon_{0}^{\mu}\epsilon_{0}^{\nu}\frac{1}{u_{L}^{2}}\text{Im}\left(h_{11}+h_{12}\right) \nonumber\\ 
&-4\beta_{0}\int_0^{\infty}\frac{dp^{0}}{\pi}p^{0}u_{\rho}\text{Im}\left(h_{16}^{T}\right)+4\beta_{0}\int_0^{\infty}\frac{dp^{0}}{\pi}p^{0}\Delta_{\mu\nu}^{L}u_{\rho}\epsilon_{0}^{\mu}\epsilon_{0}^{\nu}\text{Im}\left(h_{16}^{L}-h_{16}^{T}\right)\;,\nonumber\\
\mathcal{T}_{\rho\sigma}^{2}= & \int_0^{\infty}\frac{dp^{0}}{\pi}p^{0}\Delta_{\mu\rho}^{T}u_{\nu}\epsilon_{0}^{\mu}\epsilon_{0}^{\nu}\text{Im}\left[u_{\sigma}^{L}\left(h_{5}-h_{6}\right)+2\frac{p_{\sigma}u\cdot p}{\beta_{0}p^{2}u_{L}^{2}}\left(h_{11}+h_{12}\right)\right] \nonumber\\
&-\frac{4}{\beta_{0}}\int_0^{\infty}\frac{dp^{0}}{\pi}p^{0}p_{\rho}p_{\sigma}\text{Im}\left(h_{15}^{T}\right)+\frac{4}{\beta_{0}}\int_0^{\infty}\frac{dp^{0}}{\pi}p^{0}\Delta_{\mu\nu}^{L}p_{\rho}p_{\sigma}\epsilon_{0}^{\mu}\epsilon_{0}^{\nu}\text{Im}\left(h_{15}^{L}-h_{15}^{T}\right)\;, \nonumber\\
\mathcal{T}_{\rho\sigma}^{3}=&-\frac{2}{\beta_{0}}\int_0^{\infty}\frac{dp^{0}}{\pi}p^{0}\Delta_{\mu\rho}^{T}u_{\nu}\epsilon_{0}^{\mu}\epsilon_{0}^{\nu}\frac{1}{u_{L}^{2}}\text{Im}\left[u_{\sigma}\left(h_{11}+h_{12}\right)\right]\;,\nonumber\\
\mathcal{N}^{0}= & 2\int_0^{\infty}\frac{dp^{0}}{\pi}p^{0}n_{B}\left[3\text{Im}\left(\rho_{T}^{*}\right)+\text{Im}\left(\rho_{L}^{*}-\rho_{T}^{*}\right)\right]\;,\nonumber\\
\mathcal{N}_{\rho}^{1}= & 4\beta_{0}u_{\rho}\int_0^{\infty}\frac{dp^{0}}{\pi}p^{0}\text{Im}\left[3h_{16}^{T}+\left(h_{16}^{L}-h_{16}^{T}\right)\right]\;,\nonumber\\
\mathcal{N}_{\rho\sigma}^{2}&=\frac{4}{\beta_{0}}\int_0^{\infty}\frac{dp^{0}}{\pi}p^{0}p_{\rho}p_{\sigma}\text{Im}\left[3h_{15}^{T}+\left(h_{15}^{L}-h_{15}^{T}\right)\right]\;.
\label{eq:T_N}
\end{align}
 
\begin{table}
\begin{centering}
\begin{tabular}{|c|c|c|c|}
\hline 
 & explicit form &  & explicit form\tabularnewline
\hline 
\hline 
$h_{1}$ & $\qquad\qquad4n_{B}F_{1}\rho_{T}\text{Re}\left(i\rho_{L}^{*}\right)-iH_{1}\rho_{T}\rho_{L}^{*}\qquad\qquad$ & $h_{4}$ & $4n_{B}F_{2}\rho_{L}\text{Re}\left(i\rho_{T}^{*}\right)-iH_{2}\rho_{L}\rho_{T}^{*}$\tabularnewline
\hline 
$h_{2}$ & $\qquad\qquad4n_{B}F_{1}\rho_{L}\text{Re}\left(i\rho_{T}^{*}\right)-iH_{1}\rho_{L}\rho_{T}^{*}\qquad\qquad$ & $h_{5}$ & $4n_{B}F_{3}\rho_{T}\text{Re}\left(i\rho_{L}^{*}\right)-iH_{3}\rho_{T}\rho_{L}^{*}$\tabularnewline
\hline 
$h_{3}$ & $\qquad\qquad4n_{B}F_{2}\rho_{T}\text{Re}\left(i\rho_{L}^{*}\right)-iH_{2}\rho_{T}\rho_{L}^{*}\qquad\qquad$ & $h_{6}$ & $4n_{B}F_{3}\rho_{L}\text{Re}\left(i\rho_{T}^{*}\right)-iH_{3}\rho_{L}\rho_{T}^{*}$\tabularnewline
\hline 
 & \multicolumn{3}{c|}{explicit form}\tabularnewline
\hline 
$h_{7}$ & \multicolumn{3}{c|}{$\rho_{T}\left(\frac{1}{\rho_{L}}-\frac{1}{\rho_{T}}\right)\partial_{u\cdot p}\left(n_{B}\text{Re}\left(i\rho_{L}^{*}\right)\right)-\rho_{T}\partial_{u\cdot p}\left(\frac{1}{\rho_{T}}\right)n_{B}\text{Re}\left(i\rho_{L}^{*}-i\rho_{T}^{*}\right)+\left(\rho_{L}-\rho_{T}\right)n_{B}\text{Re}\left(\frac{i}{\rho_{L}}\right)\partial_{u\cdot p}\left(\rho_{L}^{*}\right)$}\tabularnewline
\hline 
$h_{8}$ & \multicolumn{3}{c|}{$\rho_{L}\left(\frac{1}{\rho_{L}}-\frac{1}{\rho_{T}}\right)\partial_{u\cdot p}\left(n_{B}\text{Re}\left(i\rho_{T}^{*}\right)\right)-\rho_{L}\partial_{u\cdot p}\left(\frac{1}{\rho_{L}}\right)n_{B}\text{Re}\left(i\rho_{L}^{*}-i\rho_{T}^{*}\right)+\left(\rho_{L}-\rho_{T}\right)n_{B}\text{Re}\left(\frac{i}{\rho_{T}}\right)\partial_{u\cdot p}\left(\rho_{T}^{*}\right)$}\tabularnewline
\hline 
$h_{9}$ & \multicolumn{3}{c|}{$\rho_{T}\left(\frac{1}{\rho_{L}}-\frac{1}{\rho_{T}}\right)\partial_{p^{2}}\left(n_{B}\text{Re}\left(i\rho_{L}^{*}\right)\right)-\rho_{T}\partial_{p^{2}}\left(\frac{1}{\rho_{T}}\right)n_{B}\text{Re}\left(i\rho_{L}^{*}-i\rho_{T}^{*}\right)+\left(\rho_{L}-\rho_{T}\right)n_{B}\text{Re}\left(\frac{i}{\rho_{L}}\right)\partial_{p^{2}}\left(\rho_{L}^{*}\right)$}\tabularnewline
\hline 
$h_{10}$ & \multicolumn{3}{c|}{$\rho_{L}\left(\frac{1}{\rho_{L}}-\frac{1}{\rho_{T}}\right)\partial_{p^{2}}\left(n_{B}\text{Re}\left(i\rho_{T}^{*}\right)\right)-\rho_{L}\partial_{p^{2}}\left(\frac{1}{\rho_{L}}\right)n_{B}\text{Re}\left(i\rho_{L}^{*}-i\rho_{T}^{*}\right)+\left(\rho_{L}-\rho_{T}\right)n_{B}\text{Re}\left(\frac{i}{\rho_{T}}\right)\partial_{p^{2}}\left(\rho_{T}^{*}\right)$}\tabularnewline
\hline 
$h_{11}$ & \multicolumn{3}{c|}{$\rho_{T}\left(\frac{1}{\rho_{L}}-\frac{1}{\rho_{T}}\right)\partial_{u\cdot p}\left(n_{B}\text{Re}\left(i\rho_{L}^{*}\right)\right)-\rho_{T}\partial_{u\cdot p}\left(\frac{1}{\rho_{T}}\right)n_{B}\text{Re}\left(i\rho_{L}^{*}-i\rho_{T}^{*}\right)-\partial_{u\cdot p}\left(\rho_{T}\right)n_{B}\text{Re}\left(\frac{i}{\rho_{T}}\right)\left(\rho_{L}^{*}-\rho_{T}^{*}\right)$}\tabularnewline
\hline 
$h_{12}$ & \multicolumn{3}{c|}{$\rho_{L}\left(\frac{1}{\rho_{L}}-\frac{1}{\rho_{T}}\right)\partial_{u\cdot p}\left(n_{B}\text{Re}\left(i\rho_{T}^{*}\right)\right)-\rho_{L}\partial_{u\cdot p}\left(\frac{1}{\rho_{L}}\right)n_{B}\text{Re}\left(i\rho_{L}^{*}-i\rho_{T}^{*}\right)-\partial_{u\cdot p}\left(\rho_{L}\right)n_{B}\text{Re}\left(\frac{i}{\rho_{L}}\right)\left(\rho_{L}^{*}-\rho_{T}^{*}\right)$}\tabularnewline
\hline 
$h_{13}$ & \multicolumn{3}{c|}{$\rho_{T}\left(\frac{1}{\rho_{L}}-\frac{1}{\rho_{T}}\right)\partial_{\beta^{2}}\left(n_{B}\text{Re}\left(i\rho_{L}^{*}\right)\right)-\rho_{T}\partial_{\beta^{2}}\left(\frac{1}{\rho_{T}}\right)n_{B}\text{Re}\left(i\rho_{L}^{*}-i\rho_{T}^{*}\right)-\partial_{\beta^{2}}\left(\rho_{T}\right)n_{B}\text{Re}\left(\frac{i}{\rho_{T}}\right)\left(\rho_{L}^{*}-\rho_{T}^{*}\right)$}\tabularnewline
\hline 
$h_{14}$ & \multicolumn{3}{c|}{$\rho_{L}\left(\frac{1}{\rho_{L}}-\frac{1}{\rho_{T}}\right)\partial_{\beta^{2}}\left(n_{B}\text{Re}\left(i\rho_{T}^{*}\right)\right)-\rho_{L}\partial_{\beta^{2}}\left(\frac{1}{\rho_{L}}\right)n_{B}\text{Re}\left(i\rho_{L}^{*}-i\rho_{T}^{*}\right)-\partial_{\beta^{2}}\left(\rho_{L}\right)n_{B}\text{Re}\left(\frac{i}{\rho_{L}}\right)\left(\rho_{L}^{*}-\rho_{T}^{*}\right)$}\tabularnewline
\hline 
$h_{15}^{a}$ & \multicolumn{3}{c|}{$\rho_{a}\partial_{u\cdot p}\left(\frac{1}{\rho_{a}}\right)\partial_{p^{2}}\left(n_{B}\text{Re}\left(i\rho_{a}^{*}\right)\right)-\rho_{a}\partial_{p^{2}}\left(\frac{1}{\rho_{a}}\right)\partial_{u\cdot p}\left(n_{B}\text{Re}\left(i\rho_{a}^{*}\right)\right)+\partial_{u\cdot p}\left(\rho_{a}\right)n_{B}\text{Re}\left(\frac{i}{\rho_{a}}\right)\partial_{p^{2}}\left(\rho_{a}^{*}\right)$}\tabularnewline
\hline 
$h_{16}^{a}$ & \multicolumn{3}{c|}{$\rho_{a}\partial_{\beta^{2}}\left(\frac{1}{\rho_{a}}\right)\partial_{u\cdot p}\left(n_{B}\text{Re}\left(i\rho_{a}^{*}\right)\right)-\rho_{a}\partial_{u\cdot p}\left(\frac{1}{\rho_{a}}\right)\partial_{\beta^{2}}\left(n_{B}\text{Re}\left(i\rho_{a}^{*}\right)\right)+\partial_{\beta^{2}}\left(\rho_{a}\right)n_{B}\text{Re}\left(\frac{i}{\rho_{a}}\right)\partial_{u\cdot p}\left(\rho_{a}^{*}\right)$}\tabularnewline
\hline 
$h_{17}^{a}$ & \multicolumn{3}{c|}{$\rho_{a}\partial_{\beta^{2}}\left(\frac{1}{\rho_{a}}\right)\partial_{p^{2}}\left(n_{B}\text{Re}\left(i\rho_{a}^{*}\right)\right)-\rho_{a}\partial_{p^{2}}\left(\frac{1}{\rho_{a}}\right)\partial_{\beta^{2}}\left(n_{B}\text{Re}\left(i\rho_{a}^{*}\right)\right)+\partial_{\beta^{2}}\left(\rho_{a}\right)n_{B}\text{Re}\left(\frac{i}{\rho_{a}}\right)\partial_{p^{2}}\left(\rho_{a}^{*}\right)$}\tabularnewline
\hline 
\end{tabular}
\par\end{centering}
\caption{Coefficient functions expressed in terms of spectral functions and self-energies, the upper indices $a=T,L$ in the last three terms denotes the transverse and longitudinal components.
\label{tab:h_i}}
\end{table}

\FloatBarrier
\bibliographystyle{h-physrev}
\bibliography{citation}

\end{document}